# Multiblock variable influence on orthogonal projections (MB-VIOP) for enhanced interpretation of total, global, local and unique variations in OnPLS models

Beatriz Galindo-Prieto[a,b,c,*], Paul Geladi[d], Johan Trygg[a,e,*]

[a] Computational Life Science Cluster (CLiC), Department of Chemistry (KBC), Industrial Doctoral School (IDS), Umeå University, Umeå, Sweden

[b] Department of Engineering Cybernetics (ITK), Norwegian University of Science and Technology (NTNU), Trondheim, Norway

[c] Helen and Robert Appel Alzheimer's Disease Research Institute, Feil Family Brain and Mind Research Institute, Weill Cornell Medicine (WCM), Cornell University, New York, USA

[d] Forest Biomaterials and Technology, Swedish University of Agricultural Sciences (SLU), Umeå, Sweden

[e] Sartorius Stedim Data Analytics, Umeå, Sweden

Authors' information:

* Corresponding authors: beg4004@med.cornell.edu (B. Galindo-Prieto, PhD), johan.trygg@umu.se (J. Trygg, Prof.)

# Abstract

**Background**

For multivariate data analysis involving only two input matrices (e.g., X and Y), the previously published methods for variable influence on projection (e.g., $VIP_{OPLS}$ or $VIP_{O2PLS}$) are widely used for variable selection purposes, including (i) variable importance assessment, (ii) dimensionality reduction of big data and (iii) interpretation enhancement of PLS, OPLS and O2PLS models. For multiblock analysis, the OnPLS models find relationships among multiple data matrices (more than two blocks) by calculating latent variables; however, a method for improving the interpretation of these latent variables (model components) by assessing the importance of the input variables was not available up to now.

**Results**

A method for variable selection in multiblock analysis, called multiblock variable influence on orthogonal projections (MB-VIOP) is explained in this paper. MB-VIOP is a model based variable selection method that uses the data matrices, the scores and the normalized loadings of an OnPLS model in order to sort the input variables of more than two data matrices according to their importance for both simplification and interpretation of the total multiblock model, and also of the unique, local and global model components separately. MB-VIOP has been tested using three datasets: a synthetic four-block dataset, a real three-block omics dataset related to plant sciences, and a real six-block dataset related to the food industry.

**Conclusions**

We provide evidence for the usefulness and reliability of MB-VIOP by means of three examples (one synthetic and two real-world cases). MB-VIOP assesses in a trustable and efficient way the importance of both isolated and ranges of variables in any type of data. MB-VIOP connects the input variables of different data matrices according to their relevance for the interpretation of each latent variable, yielding enhanced interpretability for each OnPLS model component. Besides, MB-VIOP can deal with strong overlapping of types of variation, as well as with many data blocks with very different dimensionality. The ability of MB-VIOP for generating dimensionality reduced models with high interpretability makes this method ideal for big data mining, multi-omics data integration and any study that requires exploration and interpretation of large streams of data.

# 1. Background

Multivariate data analysis can involve thousands of input (manifest) variables in just one data block. These variables may contain latent information that can help (i) to extract inferences and explain phenomena and relationships that might not be obvious from the experimental results obtained in the laboratory, (ii) to get a more meaningful and visual interpretation of the data, (iii) to optimize processes in both industry and research environments, and (iv) to understand the holistic pattern in complex biological systems where different parts interact by underlying connections. Compared to the analysis of a single dataset, the analysis of a large number of datasets (blocks) implies that the number of variables and their underlying inter-connections grow very much indeed; at this point, reducing the number of variables involved in the multiblock data analysis becomes a meaningful and much needed strategy.

Interest in multiblock approaches has risen in psychology[1–3], chemistry[4–7], biology[8,9] and sensory science[10,11], among other; an interest mainly motivated by the goal of extracting the maximum useful information from two or more datasets interrelated among themselves. Early multiblock methods based on projections and latent variables, e.g. partial least squares (PLS)[12,13], allowed the analysis of a limited number (usually two or three) of data matrices, but without taking full advantage of how the data blocks were connected. Two commonly used multiblock approaches based on principal components were consensus principal component analysis (CPCA)[14,15] and hierarchical principal component analysis (HPCA)[16], whose algorithms are very similar, differing only in the normalization steps[5]. For PLS applied to multiblock analysis, it is worth mentioning hierarchical partial least squares (HPLS)[14] and multiblock partial least squares (MBPLS)[17], which are similar but with two main differences: (i) the normalization is done on different model parameters, and (ii) the regression of the **Y**-block is done on different matrices[5]. Some interesting applications of multiblock-PLS were reported by Wise and Gallagher in 1996[18], and a better understanding of the underlying patterns in latent models was attempted by Kourti *et al.*[4] using multiblock multiway PLS for analyzing batch polymerization processes in 1995. Although many different multiblock methods based in different criteria and principles can be found in the literature (e.g. regularized generalized canonical correlation analysis, RGCCA[19]), this paper will mainly keep its scope inside methods based on partial least squares regression[20–30], such as sparse partial least squares presented by Le Cao *et al.*[31] (and further implemented by Rohart *et al.*[32]). Multiblock methods based on orthogonal projections have received interest within life-sciences provided the model structure it can decompose the data blocks into; two examples of this are the multi-omics factor analysis (MOFA) presented by Argelaguet *et al.* in 2018 and the N-block orthogonal projections to latent structures (OnPLS) method presented by Löfstedt and Trygg in 2011[33]. The latter can be used to provide some input parameters for improved model interpretation using MB-VIOP. From a methodology perspective, OnPLS provides means to take full advantage of the shared and unique variations of more than two data blocks. Examples of alternative methods with different objective functions include JIVE (joint and individual variation explained)[34], GSVD (generalized singular value decomposition)[35], and msPLS (multiset sparse partial least squares path modelling)[36].

The numerous variable selection methods for multivariate analysis of one data matrix[37–46] cannot handle the complexity and the underlying patterns of a large number of datasets; therefore, data integration and multiblock variable selection methods are needed. An important consideration is to be aware of the multiset structure since the integration of multiple datasets can be performed in different ways, and different methods may have specific requirements on this aspect. For instance, OnPLS followed by MB-VIOP has a similar integration framework than the N-integration of block sparse PLS requiring the same number of samples (N) for all data matrices, whilst mint sparse PLS has a K-integration (also called P-integration in the literature) framework which requires the same number of variables (K) instead of the same number of samples[32]. Besides, some methods are more suitable for improving model interpretability, whilst other are more suitable for improving predictability; hereby, the importance of selecting the appropriate variable selection method according to the purpose of the data analysis, an example of this was shown by comparing the obtained root mean square error of prediction (RMSEP) using two different variable selection methods on the Marzipan dataset in Galindo-Prieto *et al.* 2017 [47]. The fact that variable influence on projection (VIP) approaches for OPLS ($VIP_{OPLS}$)[38], O2PLS ($VIP_{O2PLS}$)[47] and OnPLS (MB-VIOP) base their calculations on the product between the normalized loadings (p) and the sum of squares of X and Y leads to an enhanced model interpretability that other methods cannot achieve. However, if the aim of the analysis is to achieve enhanced model predictability, other methods such as sparse PLS[31] (that uses the Q2 parameter as criterion to choose the number of model components, and the root means square error of prediction criterion for evaluation of the predictive power of each Y variable between the original non penalized PLS models and the sparse PLS model) may be more suitable. We include a comparison for unsupervised multiblock variable selection using the sparse PLS method for multiblock cases (block-sPLS)[32] and MB-VIOP in the Results and Discussion section.

In addition, variable selection aiming to enhance the interpretation of latent variables containing uncorrelated (orthogonal) variation can be challenging. An example of an approach able to deal with multiple datasets is the sparse generalized canonical correlation analysis (SGCCA) for variable selection that combines RGCCA with the L1-penalty[48]; however, to deal also with orthogonalization in an analysis of multiple datasets, methods such as $VIP_{O2PLS}$ (also called O2PLS-VIP)[47], MOFA[49], or the MB-VIOP explained here are more suitable options. We include a comparison for unsupervised integrated feature selection between MOFA and MB-VIOP in the Results and Discussion section.

It is worth mentioning that for one PLS component, loadings or weights can be used for determining which variables are more influential[50], but this has limited use. There is a need for a diagnostic giving the described variable influence in a PLS model, or any of its derived orthogonal versions, using more than 1 component. All VIP diagnostics are constructed for that purpose.

A multiblock variable selection method called *multiblock variable influence on orthogonal projections* (*MB-VIOP*) for OnPLS models was developed as part of previous thesis work[51] and is now published and explained in this paper. The mathematical principles of MB-VIOP relate to those used in VIP$_{OPLS}$ (a.k.a., OPLS-VIP)[38,43] and VIP$_{O2PLS}$ (a.k.a., O2PLS-VIP)[47]. However, the cornerstone of MB-VIOP is its inter-block connectivity with emphasis on the variable influence, making MB-VIOP substantially different (i) from its two predecessors VIP$_{OPLS}$ and VIP$_{O2PLS}$ in terms of connectivity, and also (ii) from OnPLS regression[33] since the normalized OnPLS **p** loadings cannot provide by themselves a reliable and precise variable importance assessment while this is easily achieved by MB-VIOP by taking these normalized loadings as starting point for the variable importance assessment (as it will be shown in the synthetic example). MB-VIOP allows the selection of the most important variables for enhanced interpretation of OnPLS models when three or more data blocks are simultaneously modelled. It is worth mentioning that MB-VIOP is also applicable to O2PLS® models that involve only two data blocks. Furthermore, MB-VIOP provides four MB-VIOP profiles (total, global, local and unique) to help answer questions such as:

a) Total MB-VIOP profile: Which are the variables that are more relevant for the interpretation of the whole model? Which variables could be eliminated from the model in order to improve it?

b) Global MB-VIOP profile: Which variables help to interpret the variation that is common to all the data blocks involved in the model?

c) Local MB-VIOP profile: Which variables are important to interpret the variation that is common to some of (but not all) the blocks? And how do these variables connect among the data blocks to explain the information shared by them (i.e., the variation related to the same component or latent variable)?

d) Unique MB-VIOP profile: Which are the variables that contain unique information that can be only found in one specific data block? And which inferences related to the data can be elucidated from the selected variables in the unique MB-VIOP profiles?

The MB-VIOP algorithm has been tested by using three multiblock datasets, (i) a simulated four-block dataset called *SD16_235GLU*, (ii) a real three-block omics dataset here called *Hybrid Aspen*, and (iii) a real six-block industrial dataset called *Marzipan*. The three datasets are described in detail in Sections 4.5 – 4.7.

## 2. Results and discussion

The results and the discussion aim to validate the multiblock variable influence on orthogonal projections (MB-VIOP) method for its application in OnPLS models (extended interpretations related to biology or spectroscopy are out of the scope of this paper). Thus, an OnPLS model followed by an MB-VIOP variable selection will be performed in all multiblock analyses. The input variables will be sorted according to their importance for the entire multiblock model (i.e., the total variation), but also for each model component separately (i.e., the unique, the local and the global variations). Figure 1 shows the different types of variation present in a generic OnPLS model.

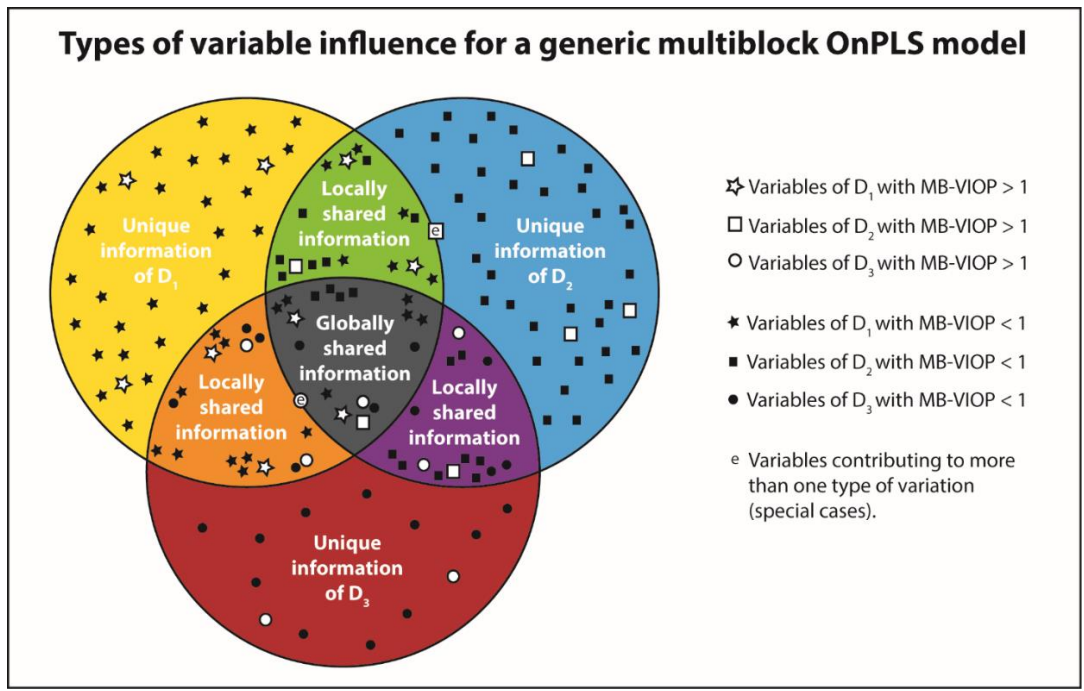

*Figure 1: Venn diagram that shows the three types of variable influences in MB-VIOP according to the type of variation (global, local or unique) that they explain. The three data blocks are represented by three big circles (yellow for $D_1$, blue for $D_2$, red for $D_3$). There are three different types of zones according to how the information is shared (i.e. globally, locally or uniquely) by the variables among the blocks. Variables that belong to $D_1$ are represented by stars, variables of $D_2$ by squares, and variables of $D_3$ by circles. Variables filled in white are important, whereas the ones filled in black are not. Variables labeled with an e are special cases. A further explanation is provided in Section 4.*

## 2.1. Description of the OnPLS models

For the synthetic four-block SD16_235GLU data, an OnPLS model was built in MATLAB. The OnPLS algorithm found two global components (in black and blue in Figure 2), three local components (in cyan, orange and green in Figure 2), and three unique components (in pink color in Figure 2); which points to a conservative, but well conducted, modelling by the OnPLS algorithm. Only two unique components included in the design of the synthetic data were not found; i.e., one unique component in block **D₁** (which represented a 14.3 % of the variation of **D₁**) and one unique component in block **D₄** (which contained a 20% of the variation of **D₄**). The rest of the variation was extracted by the model (see Table 1); the percentage of total variation explained by the model was 85.8 % for **D₁**, 100% for **D₂**, 100% for **D₃** and 80% for **D₄**.

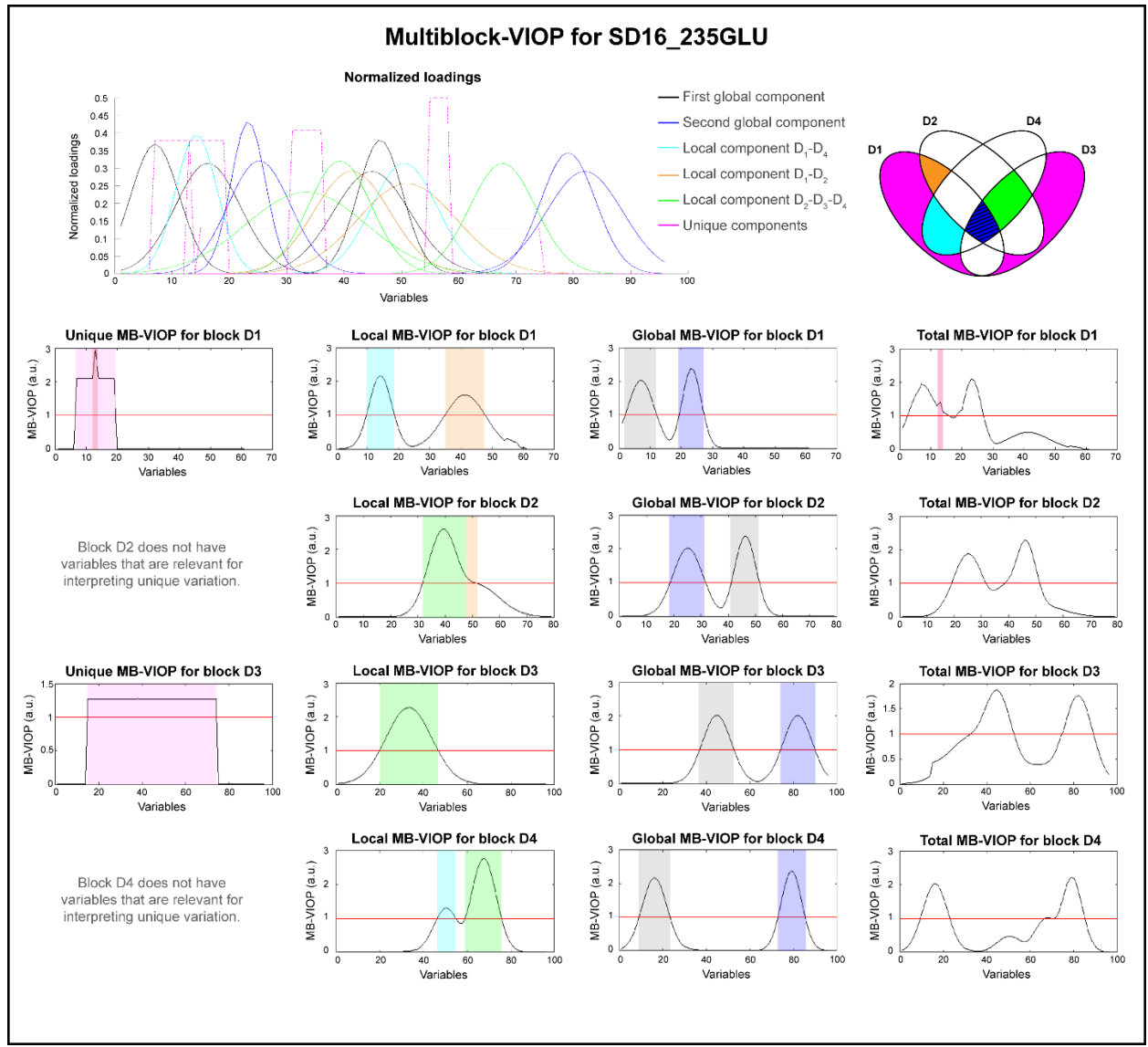

***Figure 2:*** *MB-VIOP results for the synthetic data set SD16_235GLU. An overview of the 4-block (**D₁-D₄**) system and its interactions is shown at the top right of the figure. The normalized loadings directly*

*extracted from the synthetic dataset (not from the model) are provided at the top left. For the whole figure, the color code is indicated in the legend (pink is used for unique, black and blue for global, cyan ($D_1$-$D_4$) and orange ($D_1$-$D_2$) for local information related to two-block interactions, and green for local information related to the three-block interaction ($D_2$-$D_3$-$D_4$)). The MB-VIOP plots are distributed by columns according to type of interpreted variation, and by rows according to data block. The important variables are the ones with MB-VIOP values above the red line (MB-VIOP > 1). A more detailed interpretation of the results of this figure is given in Section 2.2.*

| SD16_235GLU MODEL | | | | | | | | |
|---|---|---|---|---|---|---|---|---|
| Percentage of explained variation per data block and per component | | | | | | | | |
| Data block | $a_{g1}$ | $a_{g2}$ | $a_{l1}$ | $a_{l2}$ | $a_{l3}$ | $a_{u1}$ | $a_{u2}$ | $a_{u3}$ |
| D1 | 14.3 | 14.3 | 14.3 | 14.3 | | 14.3 | 14.3 | |
| D2 | 25.0 | 25.0 | | 25.0 | 25.0 | | | |
| D3 | 25.0 | 25.0 | | | 25.0 | | | 25.0 |
| D4 | 20.0 | 20.0 | 20.0 | | 20.0 | | | |

***Table 1:*** *Values of explained variation per data block (**D1-D4**) and per component for the OnPLS model of the SD16_235GLU dataset. Values are given as percentages (%), a stands for component, g for global, l for local, and u for unique.*

For the Marzipan data, the six data matrices were used to generate an OnPLS model, which yielded two global components and two unique components (the percentages of explained variation per component and per block are shown in Table 2). The model was able to explain almost all variation; more specifically, a 96.2 % of total variation for the NIRS1 block, a 93.8 % for the NIRS2 block, a 95.8 % for the INFRAPROVER block, a 97.0 % for the BOMEM block, a 99.9 % for the INFRATECH block and a 75.5 % for the IR block. Since all blocks are related to NIR/IR spectroscopy, it is not surprising that the OnPLS algorithm found two global components. The Marzipan data mostly has predictive (joint) variation, which is absolutely dominant over the orthogonal (unique) variation[47].

| MARZIPAN MODEL | | | | |
| --- | --- | --- | --- | --- |
| Percentage of explained variation per data block and per model component | | | | |
| Data block | $a_{g1}$ | $a_{g2}$ | $a_{u1}$ | $a_{u2}$ |
| NIRS1 | 76.3 | 11.1 | 8.8 | |
| NIRS2 | 90.5 | 3.3 | | |
| INFRAPROVER | 84.7 | 11.1 | | |
| BOMEM | 94.2 | 2.8 | | |
| INFRATECH | 99.2 | 0.7 | | |
| IR | 41.5 | 26.9 | | 7.1 |

*Table 2:* *Values of explained variation per data block and per component for the OnPLS model of the Marzipan dataset. Values are given as percentages (%), a stands for component, g for global, and u for unique.*

For the Hybrid Aspen data, an OnPLS model was built obtaining four global components, two local components (one shared between the transcript and the metabolite data, and another shared between the transcript and the protein data), and two unique components (one for the transcriptomics block, and another for the metabolomics block). The OnPLS model explained 75.0 % of the total variation for the transcriptomics data block (14738 variables), 55.0 % for the proteomics data block (3132 variables), and 58.3 % for the metabolomics data block (281 variables). The decomposition of explained variation for the different types of variation is shown in Table 3.

| HYBRID ASPEN MODEL | | | | | | | | |
| --- | --- | --- | --- | --- | --- | --- | --- | --- |
| Percentage of explained variation per data block and per component | | | | | | | | |
| Data block | $a_{g1}$ | $a_{g2}$ | $a_{g3}$ | $a_{g4}$ | $a_{l1}$ | $a_{l2}$ | $a_{u1}$ | $a_{u2}$ |
| TRANSCRIPTOMICS | 11.9 | 30.9 | 12.0 | 2.4 | 4.4 | 5.3 | 8.1 | |
| PROTEOMICS | 17.8 | 14.4 | 10.6 | 4.0 | | 8.2 | | |
| METABOLOMICS | 12.3 | 14.2 | 7.8 | 6.1 | 5.7 | | | 12.3 |

*Table 3:* *Values of explained variation per data block and per component for the OnPLS model of the Hybrid Aspen dataset. Values are given as percentages (%), a stands for component, g for global, l for local, and u for unique.*

## 2.2. Evidence of the reliability and the efficiency of MB-VIOP using synthetic data

For the variation contained in the local component that $D_1$ shares with $D_4$, MB-VIOP selected as relevant variables 10-18, represented as a peak marked in cyan in the local MB-VIOP plot for $D_1$ (Figure 2); in the same local MB-VIOP plot, variables 35-47 (marked in orange) were considered important for explaining the variation that $D_1$ shares with $D_2$. The unique MB-VIOP plot for $D_1$ pointed at variables 7-19 as the important ones for explaining the unique variation of $D_1$; interestingly, variable 13 stood out from the rest of variables.

By comparing the MB-VIOP variable importance results to the normalized loadings (Figure 2), it can be seen that the MB-VIOP method is very reliable finding the exact variables that are important for the different types of variation of $D_1$; furthermore, MB-VIOP assesses the correct proportion of importance for each variable, which cannot be achieved by the normalized loadings plot. Hence, looking at variable 13 in the normalized loadings plot, it can be seen that this variable was related to the two unique components of $D_1$ (explaining 28.6% of variation), whereas the other variables (7-12 and 14-19) linked to the unique variation of $D_1$ were only related to one of the unique components (explaining only 14.3% of the variation); however, the normalized loading plot did not highlight such an important variable (no. 13) in any way. Auspiciously, MB-VIOP highlighted the importance of variable 13 (marked in dark pink color in Figure 2) as an intense peak standing out from the crowd; this variable was also depicted in the total MB-VIOP plot for $D_1$. Therefore, the total and the unique MB-VIOP plots for $D_1$ evidence the efficiency of MB-VIOP algorithm to not lose track of any variable, even if it is a lonely variable.

The MB-VIOP results obtained for block $D_2$ are encouraging, since, even with a high overlapping of the normalized loadings (profiles), the MB-VIOP algorithm identified the variables that were relevant for each type of variation (see Figure 2).

For block **D₃**, the variables considered important in the global MB-VIOP plot (Figure 2) contributed to explain a 50% of the total variation of the OnPLS model, whilst the variables related to explain other types of variation did not overpass the 25%; therefore, the variables related to the information globally shared by all the data matrices were selected as the most important ones for the whole model, leaving out the variables related to information that was local or unique. The unique variation of **D₃** (25% of the total variation) was explained by the large range of variables 15-74. For an overview assessment of the variable importance, the total MB-VIOP plot pointed at variables 33-52 and 75-89 as the most relevant ones. Interestingly, the total MB-VIOP plot emphasizes the efficiency of MB-VIOP giving the proportionally fair importance to the variables according to the amount of information that they help to explain in the OnPLS model; the absence of the large amount of variables which were relevant for the unique variation (i.e., variables 15-74 of D₃) enlightened another achievement of the MB-VIOP algorithm: it does not matter if there is an outsize number of variables that are important for a specific type of variation, in case that their importance for interpreting/explaining variation in the whole model is not significant enough, they will not be considered relevant variables in the total MB-VIOP plot. The latter fact demonstrates that MB-VIOP properly sorts the variables according to their importance for explaining a specific type of variation.

### 2.3. Enhancement of the interpretability in an OnPLS model for the Marzipan case by using MB-VIOP

The MB-VIOP results (see Figure 3) obtained for the OnPLS model generated using the Marzipan dataset (previously described in Section 2.1) helped to better interpret the pattern of information overlapping between the six data matrices (that would be a painstaking task if it was done by using the normalized loadings provided in Figure 3). There is not significant amount of local variation in the Marzipan dataset, which explains the fact that no important variables for explaining local variation were selected by MB-VIOP. In addition, due to the extreme dominance of the joint variation over the unique variation, the MB-VIOP results for the global latent variables were very similar to the MB-VIOP results for the total variation, as can be seen by comparison of the plots in Figure 3.

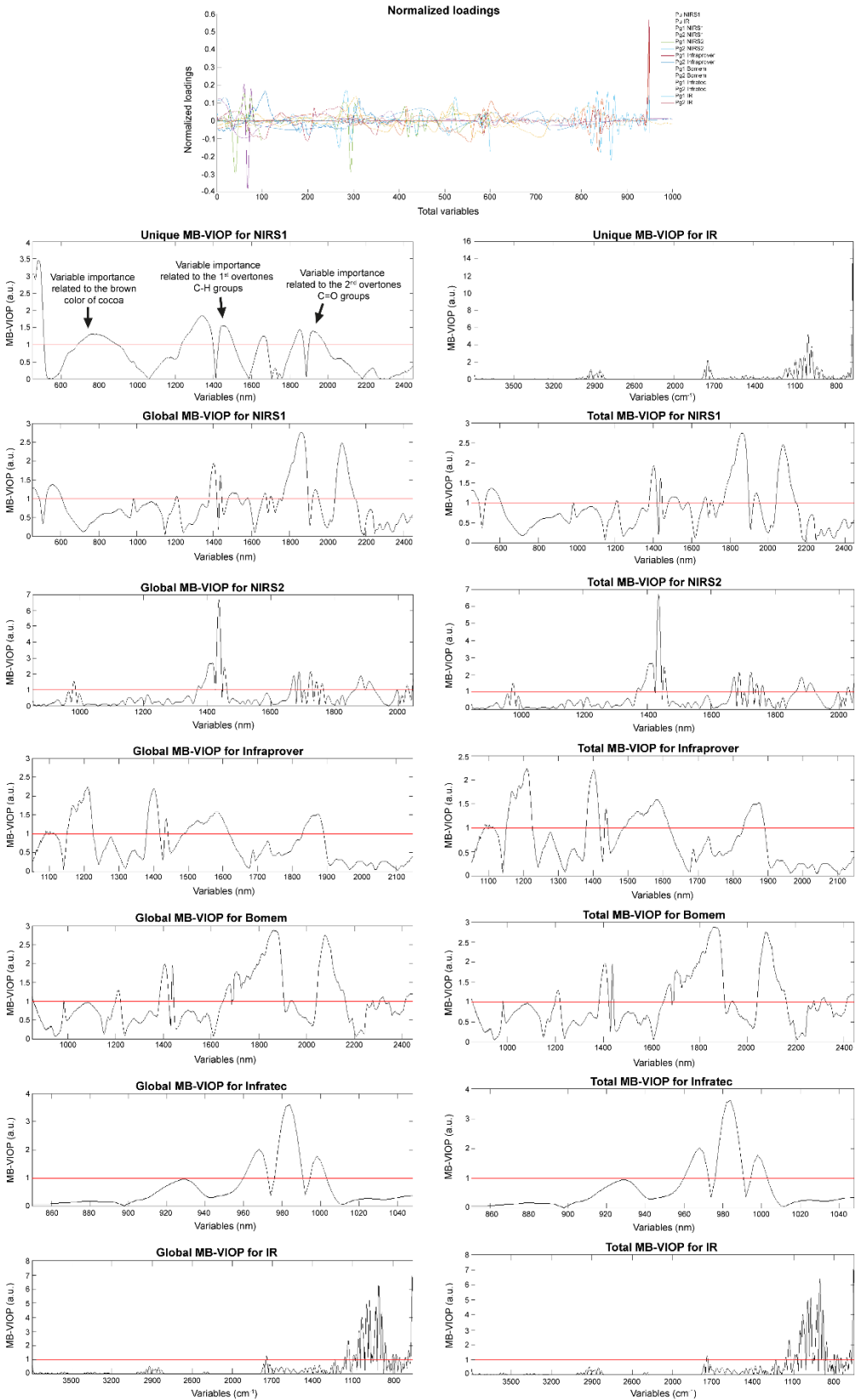

*Figure 3:* MB-VIOP results for the marzipan dataset. The normalized loadings (for all the blocks and components) obtained from the OnPLS model are provided on the top. The unique, global and total MB-VIOP plots are also provided, including the threshold line at MB-VIOP = 1. The variables determined as relevant by the MB-VIOP algorithm have been annotated in the unique MB-VIOP plot for the data block NIRS1 according to the organic compound of marzipan and/or cocoa that they help to explain.

Giving an overall look at the MB-VIOP plots of Figure 3, the manifest variables selected as relevant for the two global latent variables (global model components) seemed to relate to (i) the sugar content (majorly sucrose, but also small amounts of invert sugar and glucose syrup), and (ii) the almonds and apricot kernels. The unique MB-VIOP plots were related to special and unique characteristics of some marzipan samples and/or some spectrometers, as it will be explained in this section.

Block NIRS1 contains measurements done using an instrument that was able to cover, not only the NIR region, but also the visual light range (400-800 nm). Thanks to this, differences in color could be detected for the marzipan samples. Interestingly, MB-VIOP determined that some variables corresponding to the range between 450 and 800 nm (visual light region) were relevant for explaining variation only detectable in NIRS1 (i.e., unique for this data block). These important variables relate to the cocoa that was added to some marzipan samples (they had a more brownish color). Besides, by looking at the whole unique MB-VIOP plot (from 450 to 2448 nm) in Figure 3, it can be seen that, aside from the variables with high MB-VIOP values detected in the visual light range, there were also important variables located at 1232-1396 nm, 1428-1506 nm, 1638-1682 nm, 1818-1872 nm, and 1902-1986 nm. The cocoa NIR spectrum has been described in the literature[52], thus by matching of some of the important wavelengths found by MB-VIOP and the known composition of the cocoa, it is possible to realize the enhanced and easier model interpretation achieved by using MB-VIOP (which is not possible by using the OnPLS model loadings provided in Figure 3). The wavelengths at 1478-1506 nm are important to uncover the OnPLS model variation related to the first overtones of the C-H groups of the cocoa, and variables at 1902-1986 nm explain the variation related to the second overtones of the C=O groups of the cocoa (see Figure 3).

The Infratec MB-VIOP revealed three clear regions of important variables located at 960-972 nm, 978-990 nm and 996-1002 nm (see MB-VIOP plots for Infratec in Figure 3). These variables are selected as relevant by the MB-VIOP algorithm because they are related to the carbohydrates, proteins, water and lipids (i.e., the second overtones of O-H and N-H stretching vibrations, and the third overtones of C-H stretching vibrations). These substances are common to all the marzipan samples, which explains that these wavelengths (variables) were highlighted in the global MB-VIOP plot. It is worth noticing that these three wavelength regions can be also seen (albeit not so clearly) in the MB-VIOP plots of NIRS2.

As in the $VIP_{O2PLS}$ analysis of Marzipan data published in 2017[47], the multiblock model generated for the VIP analysis is only between spectra, not between spectra and concentrations; which can be unusual, but also useful either for technical reasons (e.g., to compare spectrometers) or for spectroscopic reasons (e.g., to see the correspondence between bands in IR and bands in NIR – overtones –). The MB-VIOP plots for NIRS1 and Bomem (Figure 3) were very similar because of the characteristics that the NIR spectrometers had in common, however MB-VIOP found some differences in the variable importance that could (maybe) be attributable to the different optical principles of the two instruments (dispersive scanning for the NIRS1, and FT interferometer for the Bomem). On the other hand, the IR data block contained relevant variables (wavenumbers) that explained information that is unique for this block, due to the differences in type of spectroscopy (IR/NIR) and instrumentation (spectrometer components).

Some very intense peaks in the MB-VIOP plots correspond to variables that are important for some major marzipan compounds. For example, the peak around 1440 nm in the MB-VIOP plot for NIRS2 could be related to the O-H bonds, and the peak around 2100 nm in the MB-VIOP plot for Bomem could relate to the protein amino acids.

### 2.4. Selection of the most relevant variables in systems biology multiblock analysis for enhanced model interpretation and dimensionality reduction.

For the Hybrid Aspen data, the variables were sorted by importance using MB-VIOP, and afterwards, this information was used for achievement of enhanced interpretability (higher percentage of explained model variation) and reduced model dimensions (less variables). The purpose was not only to validate MB-VIOP as a method for variable importance sorting, but also for multiblock variable selection. To this end, two MB-VIOP variable selections (both of them from the original model, i.e. not sequentially done) were carried out, one choosing the variables with MB-VIOP values over the default threshold (MB-VIOP $\geq 1$), and another variable selection with a more conservative criterion (i.e., MB-VIOP $\geq 0.5$). Afterwards, two new OnPLS models were generated using only the variables selected by MB-VIOP; the number of variables used in the original and the two new reduced multiblock models, as well as the percentages of total explained variation, are summarized in Table 4. We want to emphasize that the MB-VIOP profile used for selecting the variables was the total MB-VIOP because the goal was to improve the total model interpretation without focusing on any concrete part of the model. Nevertheless, it would be possible to select the variables that are more convenient for improving the interpretation of a specific type of variation (e.g., the local variation) by using its corresponding MB-VIOP profile (e.g., the local MB-VIOP) and building a new model with this selected subset of variables; hereby, MB-VIOP is a variable selection method *à la carte* according to the part of the model (total, global, local or unique) targeted to be improved. In order to show possible sensitivity differences among MB-VIOP profiles due to threshold choice (i.e., MB-VIOP ≥ 1 or MB-VIOP ≥ 0.5), the number of selected variables is shown in Table S1 in the Supporting Information and as bar plots in Figure 4 for each type of variation and each threshold choice. From Figure 4, it does not seem to exist significant differences between total and global profiles in relation to the number of selected variables. However, the number of variables selected when using the threshold MB-VIOP ≥ 1 (blue bars in Fig. 4) was clearly lower than when using the threshold MB-VIOP ≥ 0.5 (green bars in Fig. 4). For the unique variance, the reduction of number of selected variables using MB-VIOP ≥ 0.5 was substantially more significant than for the joint variation types.

| Data | OnPLS models | Number of variables used | Explained total variation (%) |
|---|---|---|---|
| **TRANSCRIPT** | Original | 14738 | 75.0 |
| | Total MB-VIOP ≥ 0.5 | 13127 | 80.1 |
| | Total MB-VIOP ≥ 1.0 | 4452 | 85.2 |
| **PROTEIN** | Original | 3132 | 55.0 |
| | Total MB-VIOP ≥ 0.5 | 2186 | 67.3 |
| | Total MB-VIOP ≥ 1.0 | 683 | 71.6 |
| **METABOLITE** | Original | 281 | 58.3 |
| | Total MB-VIOP ≥ 0.5 | 232 | 65.5 |
| | Total MB-VIOP ≥ 1.0 | 81 | 76.2 |

**Table 4:** *Summary of the number of variables used for the OnPLS models (the original and the two reduced models) and the percentages of explained total variation for the Hybrid Aspen data. The information has been distributed in three areas according to data block (transcriptomics, proteomics and metabolomics), and each area is divided in three rows: one for the original model, one for the reduced model using the variables with total MB-VIOP ≥ 0.5, and one for the reduced model using the variables with total MB-VIOP ≥ 1.*

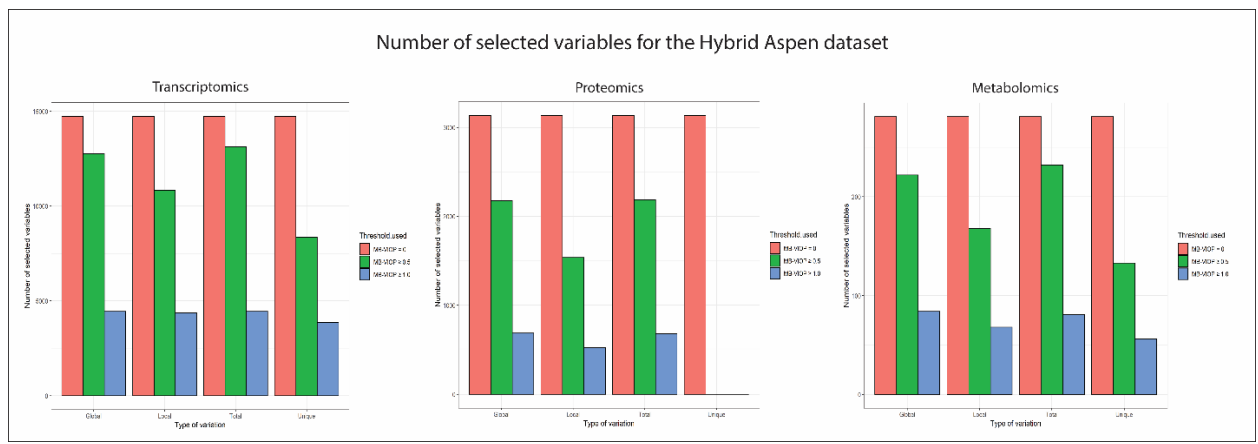

*Figure 4:* *Three plots corresponding to each Hybrid Aspen dataset grouped by type of variation. The number of variables before variable selection is represented in red, the number of variables after MB-VIOP ≥ 0.5 selection is represented in green, and the number of variables after MB-VIOP ≥ 1 selection is represented in blue.*

The blocks of the original OnPLS model contained 14738 microarray elements (variables of the transcriptomics data block) that explained the 75.0% of total variation, 3132 extracted chromatographic peaks (variables of the proteomics data block) that explained the 55.0% of total variation, and 281 extracted chromatographic peaks (variables of the metabolomics data block) that explained the 58.3% of total variation. After performing a conservative (i.e., with threshold at 0.5 a.u.) MB-VIOP selection of variables, a subset of variables was used for building a new multiblock model obtaining an increase of model interpretability; as shown in Table 4, 13127 variables from the transcriptomics data explained the 80.1% of total variation, 2186 variables from the proteomics data explained the 67.3%, and 232 variables from the metabolomics data explained the 65.5%. The second new multiblock model with reduced dimensions (using MB-VIOP ≥ 1 as criterion for selecting the subset of variables) had substantially less variables (approximately, 1/3 of the original ones) and, at the same time, increased the interpretability (measured as percentage of explained total variation in Table 4); more specifically, only 4452 transcript variables were needed to explain the 85.2% of total variation, 683 protein variables explained the 71.6%, and 81 metabolite variables the 76.2%. Due to the latter improvement, a deep exploration of the forty most important variables of each block, for interpreting the total multiblock model, was carried out. The identification of these variables is provided in Table S2 (Supporting Information) for each block.

The variables with global MB-VIOP values above the threshold (Table S3) are important for explaining the variation related to common characteristics of the growth processes of the plants, as well as both the genotype and the internode effects (common to all data blocks). Some of the most important variables to explain this latent information were PU07944 from the transcript data, the protein variables 966 and 1071, and Win022_C04 from the metabolite data.

MB-VIOP determined that the PU06931 was the most important microarray element for explaining the locally joint information, related to lignin biosynthesis, between the transcript and the protein data, with a local MB-VIOP value of 8.05 a.u. (Table S4), followed by PU07326 and PU06434; whilst for explaining the locally shared information with the metabolite data, the most important microarray elements were PU00630 (4.50 a.u.), PU03044 and PU22639. Connecting to, variable 966 (local MB-VIOP value equal to 9.76 a.u.), followed by variables 2121 and 1115, were the most important protein variables for explaining the variation locally shared with the transcriptomics block. In the metabolite space, variable Win031_C01 (5.39 a.u.), followed by Win021_C05 and Win034_C06, were selected as the most relevant metabolite variables for explaining the local variation shared with the transcript data.

The housekeeping-like events, and the differences between the instrumentation used to characterize the data in the three different platforms, were uncovered by the variables listed in Table S5 (i.e., the variables with higher values of unique MB-VIOP).

In order to explore the possibility of finding variables that could explain more than one type of variation (i.e., the special cases illustrated in Figure 1), it is worth comparing the tables and plots for the unique, local and global MB-VIOP values. For example, in this biological case, the variable Win021_C05 of the metabolomics data block helps to explain variation that is globally shared by all the data blocks, and also contributes to explain variation that is locally shared only between the metabolomics and the transcriptomics data blocks. Therefore, one variable can contain information related to more than one type of variation, and MB-VIOP is able to detect and distinguish this feature.

## 2.5. Comparison of MB-VIOP to MOFA and block-sPLS

Two unsupervised variable selection methods, i.e. block sparse partial least squares (block-sPLS) and multi-omics factor analysis (MOFA), have been compared to multiblock variable influence on orthogonal projections (MB-VIOP). All three methods have been run in symmetric mode, i.e. giving the same importance to all data blocks and considering all of them as descriptor matrices. The results have been evaluated and we present the highlighted remarks of the comparison in this section. Further details about the procedures and calculations are described in Section 4.3.

### 2.5.1. MB-VIOP and MOFA comparison for synthetic data and real omics data

In order to compare the performance of MB-VIOP and MOFA, an 8-component MOFA model was generated yielding a percentage of total explained variation of 54.5% for **D1**, 100% for **D2**, 100% for **D3** and 80% for **D4**; i.e., similar to the percentage of total explained variation obtained by MB-VIOP (85.8 % for $D_1$, 100% for $D_2$, 100% for $D_3$ and 80% for $D_4$). The distribution of the model components had similarities and differences in relation to the one obtained by MB-VIOP. Whilst MB-VIOP found two global components and three local components as expected from the design of the synthetic data, MOFA found 3 global components and three local components (see Figure S1 in the Supporting Information). For the local variation, both methods found the local components shared by D2-D3-D4 and D1-D2, but yielded different local assessments for the other latent variables. There were also differences in the discovering of the unique components; however, both methods found a unique component for D1. In general, it seems that MB-VIOP assessed better the explained variation per model component than MOFA.

Interestingly, the results of the variable selection performed by MOFA shared many similarities with MB-VIOP. When looking at the absolute MOFA loadings for the first global component, most of the variables selected by MOFA for the four data blocks were the same variables selected by MB-VIOP (marked in purple in Figure 2). The second and third components of MOFA contained a mix in the selection of the variables that seemed to partially match the variables selected by MB-VIOP for the second global component (marked in grey in Figure 2). There was also similarity in the selected variables from both methods when looking at the explained local variation, e.g. the same variables were selected as important in the absolute loadings assessment for the fourth component of MOFA and the local D1-D2 component of MB-VIOP (marked in orange in Figure 2). The evaluation of the variable selection for the unique components found by both methods, i.e. for the unique components of D1 (in pink in Figure 2), also showed a similar variable importance assessment; however, MOFA did not highlight variable 13 that helps to explain two unique components (as explained in Section 2.2) over the variables that were only helping to interpret one unique component. As an example of how the assessment has been visualized in MOFA, the absolute loading plot from MOFA for the latter example has been included as Figure S2 in the Supporting Information.

For the Hybrid Aspen case, MOFA yielded 8 model components (see Figure S3 in the Supporting Information). The total variation explained by the model was 24.6% for metabolomics, 29.5% for proteomics and 69.2% for transcriptomics. The MOFA algorithm found two global components and two unique components for the transcriptomics and the proteomics data. It also uncovered local variation shared by the transcriptomics and the metabolomics data. However, the components distribution seems difficult to assess by looking at Figure S3 due to the low values of the R2 parameter for some cases.

The variable importance assessment performed using MOFA shared some similarities with the one performed using MB-VIOP. For instance, the metabolites ranked as the most important ones in the MOFA

model (e.g. Win022_C04, Win020_C03, Win009_C09, Win034_C06, Win031_C01 or Win021_C05) were selected as important top variables to explain global variation in both MB-VIOP (Table S3 and Section 2.4) and MOFA (Figures S4-S5). The variable selection for the transcripts and the proteins was also consistent for both MB-VIOP and MOFA; e.g. top selected transcripts for explaining the unique variation in MB-VIOP (such as PU27903 or PU28218) were also determined as important by MOFA, and proteins such as 847 or 270 were also selected in both methods. For the total models, the same 2239 transcripts, 175 proteins and 32 metabolites were selected as important features by both methods.

### 2.5.2. MB-VIOP and block-sPLS comparison for the Hybrid Aspen data

For the comparison between the MB-VIOP and the block-sPLS methods, the number of variables used in the original and reduced models and the total explained variation are summarized in Tables 4-5. Both methods, as specified in Section 4.3, used similar specifications (such as the number of components for explaining the predictive variation or the constraint/penalization degree). The percentages of explained variation obtained by the block-sPLS algorithm were inferior to the ones obtained by MB-VIOP. MB-VIOP was able to explain more total variance than block-sPLS. Furthermore, when generating the models with a reduced number of variables, MB-VIOP improved the percentage of explained variation by using only the subset of MB-VIOP selected variables for the new models instead of all original variables. On the contrary, the reduced models generated by block-sPLS explained less variance than the original block-sPLS model.

| Data | Block-sPLS models | Number of variables used | Explained total variation (%) |
|---|---|---|---|
| **TRANSCRIPT** | Original block-sPLS | 14738 | 68.0 |
| | Block-sPLS comparable to MB-VIOPtot ≥ 0.5 model | 13151 | 68.0 |
| | Block-sPLS comparable to MB-VIOPtot ≥ 1.0 model | 4483 | 66.0 |
| **PROTEIN** | Original block-sPLS | 3132 | 50.0 |
| | Block-sPLS comparable to MB-VIOPtot ≥ 0.5 model | 2201 | 50.0 |
| | Block-sPLS comparable to MB-VIOPtot ≥ 1.0 model | 685 | 48.0 |
| **METABOLITE** | Original block-sPLS | 281 | 54.0 |
| | Block-sPLS comparable to MB-VIOPtot ≥ 0.5 model | 236 | 54.0 |
| | Block-sPLS comparable to MB-VIOPtot ≥ 1.0 model | 77 | 52.0 |

*Table 5:* *Summary of the number of variables used for the block-sPLS models (the original and the two reduced models) and the percentages of explained total variation for the Hybrid Aspen data. The information has been distributed in three areas according to data block (transcriptomics, proteomics and metabolomics), and each area is divided in three rows: one for the original model, one for the reduced model using a constraint degree similar to the total MB-VIOP ≥ 0.5, and one for the reduced model using a constraint degree similar to the total MB-VIOP ≥ 1.*

The overlap between the selected variables by MB-VIOP and block-sPLS was assessed. For the moderately constrained (threshold of 0.5 a.u.) reduced MB-VIOP and block-sPLS models, the same 4257 transcripts, 559 proteins, and 75 metabolites, were selected by both methods as important. For the normally constrained (threshold of 1.0 a.u.) reduced MB-VIOP and block-sPLS models, the same 2053 transcripts, 207 proteins, and 33 metabolites, were selected by both methods as important. Considering the total number of variables selected by both methods (see Tables 4-5), this seems a good overlap for the variable selection performed using MB-VIOP and block-sPLS. Besides, some variables mentioned in Section 2.4 were selected by both methods as important for interpreting the joint variation. For example, both MB-VIOP and block-sPLS selected Win022_C04 as the most important variable in the metabolomics data, and proteins such as 1071, or transcripts such as PU07944, we selected for the proteomics and the transcriptomics data respectively.

## 3. Conclusions

A novel multiblock variable selection method, called *multiblock variable influence on orthogonal projections (MB-VIOP),* has been tested and validated here. Evidence of its reliability, efficiency and usefulness have been shown. MB-VIOP can assess in a reliable and efficient way the importance of both isolated and ranges of variables in any type of data. Furthermore, MB-VIOP can deal with strong overlapping of types of variation, as well as with many data blocks with very different dimensionality. In addition, MB-VIOP connects the variables of different data matrices according to their relevance for the data interpretation of each latent variable (component) of an OnPLS model.

MB-VIOP also takes advantage of the full symmetry of the OnPLS model, which points at some advantages over the combination of sequential multiblock modelling techniques and variable selection methods. In sequential multiblock regression, even if the parameters keep the information of all parts of the sequence (i.e., other blocks of the multiblock dataset), the sequential approach only allows the weighting of the variables in a unique path (sequence) previously established, without any symmetry. Thus, the possibility of taking into account shared influences of the variables in other combinations, not considered by the pre-established path, is missing. MB-VIOP uses the symmetry of OnPLS for establishing fairer relationships/influences between variables of different blocks iterating over all components and all blocks, i.e. considering all combinations. In addition, it is worth emphasizing the ability of $VIP_{OPLS}$[38], $VIP_{O2PLS}$[47] and MB-VIOP to uncover the variables that are important for the uncorrelated (orthogonal) variation. However, for enhanced model interpretability, the synthetic example (Section 2.2) has shown how MB-VIOP surpasses any try of variable importance assessment done by means of OnPLS **p** loadings. More specifically, MB-VIOP provides a correctly proportionated importance assessment of the variables, even when the profiles are affected by high overlapping or when there is an outsizing number of variables related to a specific type of variation, assessment that cannot be achieved by the normalized OnPLS loadings.

MB-VIOP has been compared to block-sPLS and MOFA multiblock methods. Even if the comparisons are limited by the component distribution assessed by each method, the modelling and variable selection performed led to interesting conclusions. In relation to the modelling, MB-VIOP explained a higher percentage of total variation than MOFA and block-sPLS. For the feature selection, when using synthetic data, the variables selected by MB-VIOP and MOFA seemed to be consistent; however, when using real omics data, even if some of the most important variables were selected in both methods, differences in the final sorting seemed to rise when the values of the weights of the ranked variables were too adjusted. The overlapping of selected variables between block-sPLS and MB-VIOP, and MOFA and MB-VIOP, were both significant, consistent, and similar in number of variables. It is also worth mentioning, that MB-VIOP was able to keep the proportionality in the variable importance assessment (e.g., showed as a peak variable 13 of the synthetic data because of explaining more variation than the other variables); however, MOFA did not keep this proportionality as explained in the Results section.

Nevertheless, it is interesting to compare the results for the Marzipan example obtained here with the ones obtained in 2017[47], for the NIRS2 and the IR data blocks, using an O2PLS model and the $VIP_{O2PLS}$ variable selection method. As expected, the importance assessments are very similar. However, the absence of the other four data blocks in the $VIP_{O2PLS}$ variable selection[47] made the establishment of a clear relationship between the variables of the two present blocks and the variables of the four absent blocks totally impossible, which led to classify those variables as containers of orthogonal variation; however, when the variable assessment was performed in a six-block multiblock analysis with MB-VIOP, the same variables were selected as relevant for explaining variation shared between NIRS2 and the other data blocks (e.g., variables around 1200 nm, 1400 nm and 1800 nm). Hereby, when using all the blocks in a full multiblock system, the assessment was improved in relation to the two-block combination analysis.

MB-VIOP was able to reduce the number of variables of an OnPLS model (in a third for the Hybrid Aspen example) and, at the same time, increase the model interpretability. Besides, it has been shown that MB-VIOP is a variable selection method *à la carte* for OnPLS models that allows to target a concrete type of variation (global, local or unique), or, if desired, target the total model, for afterwards building a stronger reduced OnPLS model with better interpretability than the original model.

The above achievements entail valuable advantages for industry and research groups (e.g., time optimization, fast and reliable variable selection, or enhanced interpretation in multiblock analysis). We envisage the use of MB-VIOP in fields like chemistry, biology, medicine, psychology, economy, physics, cybernetics, and engineering, inter alia. Since $VIP_{OPLS}$[38] can be applied to both OPLS® and PLS models, it is expected by the authors that MB-VIOP could be successfully applied not only to OnPLS models but also to multiblock PLS (e.g., MBPLS and HPLS models). This should lead to a more reliable and accurate variable sorting/selection in the MBPLS analysis than using other methods because of the more efficient and detailed weighting of the variables (especially due to the further connectivity ability, and the use of not only the amount of variation in Y explained by the model -SSY- but also the explained amount of variation in X -SSX-) of MB-VIOP compared to PLS-VIP ($VIP_{PLS}$) method applied to multiblock analysis. The verification of the latter hypotheses is part of future work.

## 4. Methods

### 4.1. General notation

Scalars are written using italic characters (e.g. *h*, and *H*), vectors are typed in bold lower-case characters (e.g. **h**), and matrices are defined as bold upper-case characters (e.g. **H**). When necessary, the dimensions of the matrices are specified by the subscript *r x c*, where *r* is the number of rows and *c* is the number of columns. Transposed matrices are marked with the superscript T. The symbol ○ indicates a Hadamard power or product. Matrix elements are represented by the corresponding matrix italic lower-case character adding as subscripts the row and the column where they are located (e.g., for an **H** matrix, an element located in row *i* and column *k* would be indicated as $h_{ik}$). Model components are represented by *a*. Subscripts *g, l* and *u* stand for *global*, *local* and *unique* respectively. The units *a.u.* stand for *arbitrary units* for the MB-VIOP values. Notation referring to specific cases is explained *insitu*.

### 4.2. Determination of the variable importance in OnPLS models

MB-VIOP is a model based variable selection method that uses a number *n* of preprocessed data matrices (**D**), and the scores (**t**) and the normalized loadings (**p**) from an OnPLS model. The Hadamard products of the normalized loadings (denoted as $\mathbf{p}^{\circ 2}$, i.e. **p** ○ **p**) are computed, and afterwards, they are multiplied by the ratio between the variation explained by the corresponding model component and the cumulated variation. The latter sum of squares (SS) ratio helps to assess the variable importance focusing on interpretability, i.e. the SS ratio helps to know which variables are more helpful to explain the maximum amount of variation. The scores are used for the calculation of the residuals prior to computation of the sum of squares. The MB-VIOP values, which will conform the MB-VIOP vectors, are obtained by iterative calculations among both the components (latent variables) and the data matrices, with specific combinations according to the type of variation. As final step, the square root is taken, and a normalization is performed by applying the Euclidean norm (2-norm) and multiplying by the number of manifest variables raised to the ½ power. The latter explanation is the general procedure for all types of variation (see Figure 1), details and specifications are provided below. We also describe the calculations, equations (for the unique, the local, the global, and the total variations), and how to interpret the results provided by the MB-VIOP algorithm, in the subsequent sections.

### 4.2.1. Threshold of MB-VIOP values for importance assessment

The threshold for importance assessment according to the MB-VIOP values is similar to $VIP_{OPLS}$[38] and $VIP_{O2PLS}$[47] cases. Generally, variables with MB-VIOP values higher than 1 are considered important for the model interpretation, whereas variables with MB-VIOP values below 1 could be considered irrelevant. Since the sum of squares of all MB-VIOP values is equal to the number of manifest variables of the respective data matrix, the average MB-VIOP is equal to 1; therefore, if all variables would have the same contribution to the OnPLS model, they would have MB-VIOP values equal to 1. The threshold is represented in all plots by a red horizontal line at MB-VIOP = 1 for fast visual assessment. However, since this is a data-driven methodology, there can be special cases that justify the use of other threshold values according to either the goal of the variable selection or the demand level of dimensionality reduction, as shown in Section 2.4.

### 4.2.2. Calculation of MB-VIOP for the unique components

The first computation performed in the algorithm is the unique MB-VIOP (Equation 1), which allows to assess the importance of the variables related to the unique information contained in each data block. It is worth noting that the unique information contained in the unique variation (exclusive of one block, i.e. not shared with other blocks) can be elucidated focusing on a reduced subset of important variables selected by MB-VIOP without need to inspect all variables. This subset of important variables is found using Equation 1.

$$\text{MB} - \text{VIOP}_{\text{Unique }(d_i)} = \left(K_{d_i}\right)^{1/2} \cdot \left\| \sqrt{\frac{\sum_{a_u=1}^{A_u}\left(\mathbf{p}°^2_{a_u,d_i} \times SSD_{a_u,d_i}\right)}{SSD_{cum,d_i}}} \right\|_2$$

*Equation 1*

In Equation 1, $d_i$ indicates which data block we are referring to, $K$ is the number of manifest (input) variables of the data block, $A_u$ represents the total number of unique components (unique latent variables), $a_u$ indicates a specific unique component, **p** corresponds to the normalized loadings extracted from the OnPLS model, $SSD_{a_u,d_i}$ stands for sum of squares of a data block for an $a_u^{th}$ component, $SSD_{cum,d_i}$ stands for the cumulated sum of squares of a data block, and the Euclidean normalization is indicated using the subscript *2* and enclosing the normalized expression between double-line brackets.

### 4.2.3. Calculation of MB-VIOP for the local components

MB-VIOP$_{\text{Local}}$ gives values higher than 1 to those input variables that are important for explaining the variation (information) of a specific local component in an OnPLS model. The local MB-VIOP (Equation 2) is calculated iterating among all the local components, selecting the blocks that have variables locally connected (see Figure 1), and leaving out any data block that is related to either global variation or local variation linked to a different local component. Furthermore, the local part of the MB-VIOP algorithm is constrained to ignore the connection of a data block with itself, since this would increase the importance of the locally connected variables in relation to the whole model variable influence, making the weighting system unfairly favorable to the variables with locally shared information.

In Equation 2, the local MB-VIOP calculation is summarized. The calculation iterates among all the local components $A_l$, and the local MB-VIOP values for each local component are calculated considering all the combinations (direct and reverse) of the locally connected blocks, here denoted $D_{LC}$. It should be mentioned that $D_{LC}$ includes the data block $d_i$ and also the blocks connected to it ($d_{LC}$) in Equation 2. For instance, in a multiblock analysis involving four or more data blocks, if the variation of a local component is shared by three blocks, the corresponding local MB-VIOP values will be calculated using exclusively these three blocks in an iterative and exchangeable way either to provide the normalized loading (**p**) or to provide the sum of squares values (*SSD*). In the end, all three connected blocks will have contributed as both $d_i$ and $d_{LC}$ according to the specific ongoing calculation.

$$\text{MB} - \text{VIOP}_{\text{Local}\,(d_i)} = \left(K_{d_i}\right)^{1/2} \cdot \left\lVert \sqrt{\beta^{-1} \cdot \left( \frac{\sum_{a_l=1}^{A_l} \sum_{d_{LC}=1}^{D_{LC}} \left(\mathbf{p}^{\circ 2}{}_{a_l,di} \times SSD_{a_l,d_{LC}}\right)}{SSD_{cum,d_{LC}}} \right)} \right\rVert_2$$

*Equation 2*

The iterative computation of the local MB-VIOP is condensed in Equation 2, where $A_l$ represents the total number of local components, $a_l$ stands for a specific local component, $\beta$ (beta) represents the connectivity degree, $SSD_{a_l,d_{LC}}$ stands for sum of squares explained by an $a_l^{th}$ component for a data block $d_{LC}$, $SSD_{cum,d_{LC}}$ is the cumulated sum of squares of the data block $d_{LC}$. The rest of nomenclature is analogous to Section 4.2.2.

The connectivity degree $\beta$ is based on the number of local connections, which makes MB-VIOP different from VIP$_{\text{O2PLS}}$, since the latter uses the number of local components. It is worth noting that in VIP$_{\text{O2PLS}}$ the number of local components will always be equal to the number of local connections among blocks since there are only two-block connections (since O2PLS cannot handle more than two blocks). However, in MB-VIOP, there can be connections among more than two blocks related to the same local component, which implies that the number of local components will not match the number of connections. Hereby, the connectivity degree is different in MB-VIOP.

### 4.2.4. Calculation of MB-VIOP for the global components

MB-VIOP$_{\text{Global}}$ pinpoints the variables that are relevant for explaining the variation (information) that is shared by all the data blocks related to a specific global component (these variables would be the ones filled in white inside the grey zone of Figure 1), e.g., a common biological effect present in all data matrices.

The global MB-VIOP (Equation 3) is calculated by iterating over all the data block combinations (direct and reverse modes) and all the global components. In Equation 3, for a more intuitive explanation, $d_i$ is used as the data block to which the normalized loading of an iteration belongs, and $d_j$ as the data block to which the SSD values of an iteration belong. The blocks exchange these roles on the spot (i.e., at the exact iteration corresponding to a specific calculation); thus, all **D** data blocks are used as both $d_i$ and $d_j$, but in different moments of the global MB-VIOP computation.

$$\text{MB} - \text{VIOP}_{\text{Global}(d_i)} = (K_{d_i})^{1/2} \cdot \left\| \sqrt{\frac{\sum_{a_g=1}^{A_g} \sum_{d_j=1}^{D_j=D} \left( \mathbf{p}^{\circ 2}_{a_g, d_i} \times SSD_{a_g, d_j} \right)}{SSD_{cum, d_j}}} \right\|_2$$

*Equation 3*

In Equation 3, $A_g$ represents the total number of global components (global latent variables), $a_g$ indicates a specific global component, $SSD_{a_g,d_j}$ stands for sum of squares of an $a_g{}^{th}$ component related to a data block $d_j$, and $SSD_{cum,d_j}$ stands for the cumulated sum of squares of the data block $d_j$, and the rest of nomenclature is analogous to Equations 1 and 2.

### 4.2.5. Calculation for the total variable influence for interpreting the whole model

The overview of which variables are more relevant for the total model interpretation (i.e., considering the global, the local and the unique variations involved in the OnPLS model) is highly appreciated in industrial environments; this is achieved by MB-VIOP$_{\text{Total}}$. In the total MB-VIOP the contributions of the global, local and unique MB-VIOP vectors are joined achieving a proper weighting of all variables for the total variable influence on all projections. Equation 4 summarizes its computation.

$$\text{MB} - \text{VIOP}_{\text{Total}(d_i)} = (K_{d_i})^{1/2} \cdot$$

$$\left\| \sqrt{\left(\text{MB} - \text{VIOP}_{\text{Unique}(d_i)}\right)^2 + \left(\text{MB} - \text{VIOP}_{\text{Local}(d_i)}\right)^2 + \left(\text{MB} - \text{VIOP}_{\text{Global}(d_i)}\right)^2} \right\|_2$$

*Equation 4*

The nomenclature of Equation 4 is analogous to the nomenclature mentioned in the previous sections. As in the other cases, MB-VIOP leads to a vector which contains the MB-VIOP values for the variables of each data block (but the calculations take all blocks into consideration). As it will be explained in Section 4.2.6, the visualization by plotting the MB-VIOP vectors for each block is one of the various options.

### 4.2.6. Graphical representation of the MB-VIOP results for variable importance assessment

Equations 1-4 lead to four MB-VIOP vectors (i.e., MB-VIOP$_{Unique}$, MB-VIOP$_{Local}$, MB-VIOP$_{Global}$, MB-VIOP$_{Total}$). It is always possible to look at the numerical values of MB-VIOP for each variable of the OnPLS model to assess their importance for the data interpretation. However, this can become a very time-consuming and painstaking task. Hence, a reduced table containing only target variables and its MB-VIOP values, or a graphical representation of these MB-VIOP vectors, seem a more convenient way to present the results. The MATLAB code created for MB-VIOP allows several ways to plot the results; for this paper, block-wise plots have been chosen (even though the calculation of each MB-VIOP has involved all the data blocks because of being a multiblock variable sorting). Other graphical representations could be possible; in a case where all data blocks of the OnPLS model would contain the same manifest variables, it would be possible to make a 3D (cube) plot locating the manifest variables on the X-axis, labeling the data blocks on the Y-axis, and inserting the MB-VIOP values on the Z-axis (the vertical one); this visualization becomes ideal for matrices with the same variables (e.g., in some comparison studies), but it is not recommended when the data blocks have different variables (which is frequently the case).

In Section 2, the results were represented visualizing the MB-VIOP values for each data block (by rows in the figures), and for type of variation interpreted by the variables (by columns); thus, each column of plots separately represents the unique, the local, the global and the total MB-VIOP results (Figure 2 can be used as an example). As mentioned in Section 4.2.1, a threshold at MB-VIOP = 1 (represented by a red horizontal line) is included in each plot; variables with values above the red line are relevant for the interpretation of the type of variation corresponding to the plotted MB-VIOP. The variables of different blocks that contribute to explain the same variation (e.g., a common biological effect among data blocks, or a common feature of several instruments) are marked with the same color in all block-wise plots (see Figure 2).

### 4.3. Determination of variable importance in block-sPLS and MOFA for comparison to MB-VIOP variable selection.

#### 4.3.1 Variable importance assessment using MOFA on the SD16-365GLU and the Hybrid Aspen data

MOFA[49] performs unsupervised data integration aiming to uncover the principal sources of variation in multi-omics datasets, and, in some aspects, it can be seen as a statistical generalization of principal component analysis for omics data. MOFA infers a set of factors (model components) that contain biological or technical variation that can be either shared by multiple data matrices or unique of a specific data matrix. MOFA achieves factor-wise sparsity by identifying factors (model components), but also feature-wise sparsity by means of the variable weights.

For the synthetic data, a MOFA model was generated yielding 8 latent factors (model components). The weights were plot as shown in Figure S2 and the explained variation was calculated. For the Hybrid Aspen data, an 8-component MOFA model was generated. Due to MOFA characteristics, 314 protein variables needed to be removed because of having nearly zero variance, and the model was built with the remaining 2818 protein variables. The absolute loadings were plotted, and the explained variation of the model calculated.

#### 4.3.2. Variable importance assessment using block-sPLS

Block-sPLS[31,32] is a one-step method that combines data integration and variable selection by using partial least squares (PLS). Sparsity is achieved by applying a LASSO penalization of the PLS loading vectors when computing a singular value decomposition. The Q2 parameter is used to select the number of model components, and the root mean square error of prediction serves as criterion to evaluate the predictive power of the variables between the original (non-penalized) PLS model and the sparse PLS model. Therefore, the resulting selected variables are appropriate for prediction purposes.

In order to compare the feature selection results of MB-VIOP and block-sPLS, three 6-component block-sPLS models were generated using different constraint degrees for the Hybrid Aspen data (see Table 5). Both canonical and regression modes were tested, leading to better results when the canonical approach was used. The model was built using the canonical mode available from the mixOmics R-package that is appropriate to ensure that all data matrices are considered descriptors in a symmetric framework similar to the one used in MB-VIOP. A design matrix was set to maximize correlations among the data blocks. The resulting selected variables and the percentage of total explained variation were compared to MB-VIOP.

### 4.4. Materials and software

The code of the MB-VIOP algorithm was developed using MATLAB version R2019b (The MathWorks Inc., Natick, MA, USA). The four-block synthetic data set (*SD16_235GLU*), the block-scaling preprocesses, the OnPLS models, and the MB-VIOP results (values and plots) were also done using MATLAB (The MathWorks Inc., Natick, MA, USA). The Marzipan dataset[53] was provided by the University of Copenhagen through the website www.models.life.ku.dk/Marzipan, and preprocessed using PLS-toolbox version 8.1.1 (Eigenvector Research, Inc.). The block-sPLS analysis was performed using the mixOmics R-package version 6.8.5. The MOFA analysis was performed using the MOFA R-package version 1.6.1.

### 4.5. Synthetic dataset (four blocks)

The synthetic dataset, named *SD16_235GLU*, was created by the authors for testing and validating the MB-VIOP MATLAB code. The name of the dataset, *SD16_235GLU*, stands for synthetic data (SD) designed in 2016 for having 2 global components (G), 3 local components (L), and 5 unique components (U). The dataset is conformed of four data blocks (**D₁, D₂, D₃, D₄**) and 50 observations (samples) common to all blocks. The first block (**D₁**) contains 61 manifest variables, the second block (**D₂**) contains 79, and the third and fourth blocks (**D₃** and **D₄**) contain 96 manifest variables each one. The joint (predictive) normalized loadings (**p$_g$, p$_l$**) were created using Gaussian pure profiles, which are visualized as a bell shape in the plots; whereas the unique (orthogonal) normalized loadings (**p$_u$**) were created using unit pulse pure profiles, visualized as a rectangular step in the plots. The scores, both predictive (**t$_g$, t$_l$**) and orthogonal (**t$_u$**), were randomly generated, mean-centered, scaled to unit norm, and orthogonalized among themselves. The latent variables (components) were calculated as the individual products of scores and transposed normalized loadings (**t$_a$*p$_a$$^T$**). Finally, the four data blocks were created as the sum of global, local and unique components plus the residual matrices **R**. The noise was randomized, and its level was set to 0.1%. A generic **D**-block is described in Equation 5; where $A_g$ stands for the total number of global components, $A_l$ represents the total number of local components, and $A_u$ the total number of unique components. All blocks follow the pattern of Equation 5.

$$\mathbf{D} = \sum_{a_g}^{A_g} \mathbf{t}_{a_g} \mathbf{p}_{a_g}^T + \sum_{a_l}^{A_l} \mathbf{t}_{a_l} \mathbf{p}_{a_l}^T + \sum_{a_u}^{A_u} \mathbf{t}_{a_u} \mathbf{p}_{a_u}^T + \mathbf{R}$$

*Equation 5*

Equations 6 – 9 show the combination of components for each data matrix. To simulate a global component, the corresponding score vector (**t$_{ag}$**) was shared among all blocks; for the local components, the corresponding score vector (**t$_{al}$**) was shared among the locally connected blocks for that specific local component; and for the unique components individual scores (**t$_{au}$**) were used.

$$\mathbf{D_1} = \mathbf{t}_{a_g1} * \mathbf{p}_{a_g1(D1)}^T + \mathbf{t}_{a_g2} * \mathbf{p}_{a_g2(D1)}^T + \mathbf{t}_{a_l1} * \mathbf{p}_{a_l1(D1)}^T + \mathbf{t}_{a_l2} * \mathbf{p}_{a_l2(D1)}^T + \mathbf{t}_{a_u1} * \mathbf{p}_{a_u1(D1)}^T + \mathbf{t}_{a_u2} * \mathbf{p}_{a_u2(D1)}^T + \mathbf{t}_{a_u3} * \mathbf{p}_{a_u3(D1)}^T + \mathbf{R_1}$$

*Equation 6*



$$D_2 = t_{a_g1} * p^T_{a_g1(D2)} + t_{a_g2} * p^T_{a_g2(D2)} + t_{a_l2} p^T_{a_l2(D2)} + t_{a_l3} * p^T_{a_l3(D2)} + R_2$$

Equation 7

$$D_3 = t_{a_g1} * p^T_{a_g1(D3)} + t_{a_g2} * p^T_{a_g2(D3)} + t_{a_l3} * p^T_{a_l3(D3)} + t_{a_u4} * p^T_{a_u4(D3)} + R_3$$

Equation 8

$$D_4 = t_{a_g1} * p^T_{a_g1(D4)} + t_{a_g2} * p^T_{a_g2(D4)} + t_{a_l1} * p^T_{a_l1(D4)} + t_{a_l3} * p^T_{a_l3(D4)} + t_{a_u5} * p^T_{a_u5(D4)} + R_4$$

Equation 9

The *SD16_235GLU* was designed (i) to be exigent/difficult in relation to the five unique components when modelling, (ii) to have one local component shared by three data blocks ($D_2$, $D_3$, $D_4$), (iii) to have a local component shared by $D_1$ and $D_4$, (iv) to have a local component shared by $D_1$ and $D_2$, and (v) to have two global components shared by all data blocks. The percentage of variation per component is: 14.3% in $D_1$, 25% in $D_2$, 25% in $D_3$, and 20% in $D_4$ (thus, $D_1$ has a total of seven components, $D_2$ has four, $D_3$ also four, and $D_4$ has five).

### 4.6. Marzipan dataset (six blocks).

The Marzipan dataset consists of six data blocks obtained from the analysis of thirty-two marzipan samples, of nine different recipes, performed using six different spectrometers set-ups. The marzipan samples contained different amounts of almonds, apricot kernels, water, sucrose, invert sugar, glucose syrup, and minor contributions of additives; cocoa was added in some of the marzipan samples, giving them a distinctive brown color. The six spectrometers (including optical principles, spectral range, and other details) were described by Christensen *et al.*[53] in 2004. An additional set of measurements using an InfraAlyzer 260 spectrometer was originally considered as a seventh data block[53], but it has been excluded from this work because of not using exactly the same samples than the other six instrumental analyses.

The first data block (NIRS1) contained 1000 variables (400 - 2500 nm), and the second data block (NIRS2) had 600 variables (800 - 2100 nm); both NIRS1 and NIRS2 datasets were obtained using a NIRSystems 6500 spectrometer. The third (from an Infraprover II instrument) contained 406 variables, the fourth (from a Bomem MB 160 Diffusir) consisted of 664 variables, the fifth (from an Infratec 1255) had 100 variables, and the sixth (from a PerkinElmer System 2000) had 950 variables. Thus, the dimensions of the different data blocks varied from 100 to 1000 variables (i.e., a ten times difference between the smallest and the largest). NIRS1, Infraprover II and Bomem data blocks were preprocessed by extended multiplicative signal correction (EMSC)[54]; whilst NIRS2, Infratec and PerkinElmer data blocks were preprocessed by Savitsky-Golay differentiation ($2^{nd}$ derivative, $3^{rd}$ order, 15 points window)[55]. In addition, all data blocks were mean-centered and normalized to equal sum of squares before building the OnPLS model.

### 4.7. Metabolomics, proteomics and transcriptomics data of hybrid aspen (three blocks).

The Hybrid Aspen dataset used here, previously pretreated and analyzed in Bylesjö *et al.*[56] in 2009 and in Löfstedt *et al.*[57] in 2013, contains thirty-three samples of hybrid aspen (*Populus tremula x Populus tremuloides*) labeled according to the plant internode from where they were sampled (categories A, B, and C) and according to three different genotypes of hybrid aspen (WT, G5, and G3). The wild type (WT) played the role of reference sample. The G5 and G3 genotypes were related to the *PttMYB21a* gene, which is known to primarily affect lignin biosynthesis and plant growth characteristics. The G5 genotype contained several antisense constructs of the *PttMYB21a* gene, affecting plant growth; thus, this genotype displays a distinct phenotype with slower growth compare to the WT samples. The G3 genotype contained only one antisense construct of the *PttMYB21a* gene, displaying a similar but less distinct phenotype compared to the G5 samples. Further details are described by Bylesjö *et al.*[56].

All thirty-three samples were measured for transcript (cDNA), protein (UPLC/MS) and metabolite (GC/TOFMS) quantities[57]. As result, three data blocks were obtained: a transcript data block containing 14738 variables (microarray elements), a protein data block containing 3132 variables (extracted chromatographic peaks), and a metabolite data block containing 281 variables (extracted chromatographic peaks).

# 5. List of abbreviations

**block-sPLS:** multiblock sparse partial least squares

**CPCA:** consensus principal component analysis

**GSVD:** generalized singular value decomposition

**HPCA:** hierarchical principal component analysis

**HPLS:** hierarchical partial least squares

**JIVE:** joint and individual variation explained

**MBPLS:** multiblock partial least squares

**MB-VIOP:** multiblock variable influence on orthogonal projections

**MOFA:** multi-omics factor analysis

**msPLS:** multiset sparse partial least squares

**OPLS:** orthogonal projections to latent structures

**O2PLS:** 2-block orthogonal projections to latent structures

**OnPLS:** N-block orthogonal projections to latent structures

**PCA:** principal component analysis

**PLS:** partial least squares to latent structures

**RGCCA:** regularized generalized canonical correlation analysis

**RMSEP:** root mean square error of prediction

**SGCCA:** sparse generalized canonical correlation analysis

**sPLS:** sparse partial least squares

**SSX:** sum of squares of X

**SSY:** sum of squares of Y

**VIP:** variable influence on projection

# 6. Declarations

## 6.1. Ethics approval and consent to participate

Not applicable.

## 6.2. Consent for publication

Not applicable

## 6.3. Availability of data and materials

The Marzipan dataset analyzed in the current study is available through the website www.models.life.ku.dk/Marzipan of the University of Copenhagen. The Hybrid Aspen and the SD16_235GLU datasets are available from the authors on reasonable request.

## 6.4. Competing interests

The authors declare that they have no competing interests.

## 6.5. Funding

The authors are grateful for the financial support given by MKS Instruments AB (BGP), eSSENCE (JT), and Industrial Doctoral School (BGP), Umeå University, Sweden. In addition, part of this work was carried out during the tenure of an ERCIM "Alain Bensoussan" Fellowship Programme (BGP). The funding body did not play any roles in the design of the study and collection, the analysis, the data interpretation, or the manuscript writing.

## 6.6. Authors' contributions

For the MB-VIOP algorithm, BGP generated the theory, equations, MATLAB code, results and figures for the three datasets during her PhD under the supervision of JT and PG. BGP also generated the R codes and results for the comparisons using the MOFA and the block-sPLS methods. JT provided the OnPLS models generated using the OnPLS algorithm/code. PG advised on the theory and equations of MB-VIOP, method validation and spectroscopy interpretation. BGP wrote the manuscript draft, and PG checked it and improved it. The manuscript was revised and approved by all authors.

## 6.7. Acknowledgements

The authors want to thank the anonymous reviewers for helping to improve this paper, the Chemistry Department of Umeå University where MB-VIOP was developed as part of the PhD thesis of BGP, and the University of Copenhagen for providing the Marzipan dataset via the website www.models.life.ku.dk/Marzipan.

T., Wingsle, G. & Trygg, J. Integrated analysis of transcript, protein and metabolite data to study lignin biosynthesis in hybrid aspen. *Journal of Proteome Research* **8**, 199–210 (2009).

57. Löfstedt, T., Hoffman, D. & Trygg, J. Global, local and unique decompositions in OnPLS for multiblock data analysis. *Analytica Chimica Acta* **791**, 13–24 (2013).

# SUPPORTING INFORMATION:

**Contents:**

Tables S1-S5

Figures S1-S5

| Data | Threshold used | Number of variables for the total variation | Number of variables for the global variation | Number of variables for the local variation | Number of variables for the unique variation |
|---|---|---|---|---|---|
| TRANSCRIPT | No threshold | 14738 | 14738 | 14738 | 14738 |
| TRANSCRIPT | MB-VIOP ≥ 0.5 | 13127 | 12759 | 10824 | 8368 |
| TRANSCRIPT | MB-VIOP ≥ 1.0 | 4452 | 4451 | 4370 | 3860 |
| PROTEIN | No threshold | 3132 | 3132 | 3132 | 3132 |
| PROTEIN | MB-VIOP ≥ 0.5 | 2186 | 2175 | 1536 | 0 |
| PROTEIN | MB-VIOP ≥ 1.0 | 683 | 686 | 526 | 0 |
| METABOLITE | No threshold | 281 | 281 | 281 | 281 |
| METABOLITE | MB-VIOP ≥ 0.5 | 232 | 222 | 168 | 133 |
| METABOLITE | MB-VIOP ≥ 1.0 | 81 | 84 | 68 | 56 |

***Table S1:*** *Number of variables of each omics dataset of the Hybrid Aspen classified by variation type (total, global, local and unique) for the original data (no variable selection threshold applied), for the MB-VIOP selection using threshold ≥ 0.5, and for the MB-VIOP selection using threshold ≥ 1.0.*

| THE 120 MOST IMPORTANT VARIABLES FOR THE TOTAL MODEL (40 VARIABLES PER DATA BLOCK) | | | | | |
|---|---|---|---|---|---|
| Transcript variables | Total MB-VIOP values (a.u.) | Protein variables | Total MB-VIOP values (a.u.) | Metabolite variables | Total MB-VIOP values (a.u.) |
| PU07944 | 4,48 | 966 | 7,77 | Win022_C04 | 3,38 |
| PU27899 | 3,95 | 1071 | 6,95 | Win021_C05 | 2,94 |
| PU22639 | 3,85 | 3061 | 6,07 | Win034_C02 | 2,70 |
| PU23318 | 3,84 | 795 | 6,01 | Win025_C01 | 2,63 |
| PU22268 | 3,84 | 270 | 5,84 | Win023_C05 | 2,33 |
| PU28218 | 3,79 | 2805 | 5,68 | Win005_C05 | 2,31 |
| PU05769 | 3,75 | 2914 | 5,66 | Win034_C04 | 2,26 |
| PU28785 | 3,72 | 521 | 5,52 | Win007_C09 | 2,11 |
| PU25376 | 3,69 | 2121 | 5,46 | Win016_C04 | 2,05 |
| PU28089 | 3,68 | 29 | 5,38 | Win018_C10 | 2,01 |
| PU27903 | 3,65 | 2481 | 5,06 | Win022_C08 | 1,98 |
| PU28078 | 3,62 | 1125 | 5,06 | Win023_C09 | 1,89 |
| PU21598 | 3,56 | 1644 | 5,04 | Win013_C03 | 1,83 |
| PU25170 | 3,56 | 994 | 4,86 | Win020_C03 | 1,74 |
| PU27888 | 3,53 | 2305 | 4,67 | Win029_C03 | 1,72 |
| PU03044 | 3,52 | 193 | 4,57 | Win031_C01 | 1,70 |
| PU12408 | 3,50 | 527 | 4,56 | Win009_C09 | 1,69 |
| PU28695 | 3,49 | 164 | 4,49 | Win029_C01 | 1,68 |
| PU05015 | 3,47 | 1115 | 4,29 | Win013_C11 | 1,67 |
| PU28673 | 3,43 | 1609 | 4,11 | Win022_C03 | 1,66 |
| PU22628 | 3,43 | 3032 | 4,04 | Win018_C01 | 1,65 |
| PU20341 | 3,38 | 2181 | 3,99 | Win017_C02 | 1,62 |
| PU05133 | 3,38 | 2189 | 3,92 | Win003_C04 | 1,61 |
| PU22619 | 3,37 | 1702 | 3,91 | Win033_C02 | 1,56 |
| PU27826 | 3,35 | 1545 | 3,88 | Win015_C08 | 1,54 |
| PU26045 | 3,25 | 3111 | 3,86 | Win015_C06 | 1,52 |
| PU25150 | 3,24 | 757 | 3,74 | Win024_C04 | 1,51 |
| PU22573 | 3,23 | 206 | 3,72 | Win007_C13 | 1,48 |
| PU29385 | 3,21 | 969 | 3,62 | Win009_C03 | 1,44 |
| PU27678 | 3,17 | 287 | 3,57 | Win013_C09 | 1,42 |
| PU08985 | 3,17 | 934 | 3,56 | Win019_C03 | 1,39 |
| PU28223 | 3,16 | 2086 | 3,49 | Win026_C03 | 1,39 |
| PU10604 | 3,16 | 3118 | 3,38 | Win026_C07 | 1,38 |
| PU24177 | 3,16 | 847 | 3,31 | Win001_C02 | 1,38 |
| PU02239 | 3,13 | 2483 | 3,28 | Win030_C03 | 1,37 |
| PU23017 | 3,13 | 2153 | 3,26 | Win007_C03 | 1,34 |
| PU28528 | 3,12 | 1452 | 3,21 | Win002_C04 | 1,32 |
| PU02687 | 3,12 | 2923 | 3,19 | Win022_C10 | 1,31 |
| PU29692 | 3,12 | 317 | 3,19 | Win034_C01 | 1,31 |
| PU28093 | 3,12 | 1403 | 3,18 | Win034_C03 | 1,30 |

**Table S2:** *Identification of the forty most important variables for each block of the Hybrid Aspen data according to their relevance for the total model interpretation (of the original OnPLS model described in Section 2). The total MB-VIOP values are provided in arbitrary units (a.u.).*

| THE 120 MOST IMPORTANT VARIABLES FOR THE GLOBALLY JOINT VARIATION (40 VARIABLES PER BLOCK) | | | | | |
|---|---|---|---|---|---|
| Transcript variables | Global MB-VIOP values (a.u.) | Protein variables | Global MB-VIOP values (a.u.) | Metabolite variables | Global MB-VIOP values (a.u.) |
| PU07944 | 4,61 | 966 | 7,71 | Win022_C04 | 3,53 |
| PU27899 | 4,11 | 1071 | 7,03 | Win021_C05 | 2,97 |
| PU23318 | 3,99 | 3061 | 6,11 | Win034_C02 | 2,81 |
| PU22268 | 3,96 | 795 | 6,04 | Win025_C01 | 2,77 |
| PU05769 | 3,90 | 270 | 5,90 | Win023_C05 | 2,39 |
| PU22639 | 3,90 | 2914 | 5,73 | Win005_C05 | 2,39 |
| PU25376 | 3,84 | 2805 | 5,65 | Win034_C04 | 2,27 |
| PU28089 | 3,83 | 521 | 5,58 | Win022_C08 | 2,07 |
| PU28078 | 3,77 | 29 | 5,44 | Win018_C10 | 2,03 |
| PU25170 | 3,68 | 2121 | 5,36 | Win013_C03 | 1,85 |
| PU27888 | 3,67 | 1125 | 5,12 | Win020_C03 | 1,82 |
| PU28218 | 3,64 | 2481 | 5,11 | Win009_C09 | 1,78 |
| PU28785 | 3,64 | 1644 | 5,11 | Win016_C04 | 1,77 |
| PU12408 | 3,63 | 994 | 4,88 | Win029_C01 | 1,75 |
| PU05015 | 3,62 | 2305 | 4,74 | Win013_C11 | 1,73 |
| PU27903 | 3,56 | 193 | 4,63 | Win022_C03 | 1,72 |
| PU03044 | 3,54 | 527 | 4,54 | Win029_C03 | 1,70 |
| PU22628 | 3,53 | 164 | 4,44 | Win018_C01 | 1,70 |
| PU05133 | 3,51 | 1609 | 4,17 | Win023_C09 | 1,68 |
| PU28673 | 3,49 | 1115 | 4,13 | Win003_C04 | 1,67 |
| PU21598 | 3,48 | 3032 | 4,10 | Win017_C02 | 1,66 |
| PU27826 | 3,43 | 2181 | 4,04 | Win033_C02 | 1,64 |
| PU20341 | 3,42 | 2189 | 3,96 | Win024_C04 | 1,59 |
| PU28695 | 3,39 | 1702 | 3,96 | Win031_C01 | 1,56 |
| PU26045 | 3,38 | 1545 | 3,89 | Win007_C13 | 1,51 |
| PU25150 | 3,35 | 3111 | 3,88 | Win009_C03 | 1,47 |
| PU22573 | 3,33 | 206 | 3,77 | Win026_C03 | 1,45 |
| PU10604 | 3,26 | 757 | 3,67 | Win001_C02 | 1,45 |
| PU24177 | 3,26 | 934 | 3,61 | Win013_C09 | 1,44 |
| PU29385 | 3,26 | 2086 | 3,50 | Win026_C07 | 1,43 |
| PU22619 | 3,26 | 969 | 3,44 | Win030_C03 | 1,41 |
| PU28223 | 3,25 | 287 | 3,41 | Win007_C09 | 1,40 |
| PU29692 | 3,24 | 3118 | 3,38 | Win019_C03 | 1,39 |
| PU02687 | 3,21 | 847 | 3,35 | Win022_C10 | 1,37 |
| PU08985 | 3,21 | 2153 | 3,30 | Win034_C01 | 1,37 |
| PU02239 | 3,20 | 2483 | 3,24 | Win002_C04 | 1,35 |
| PU27678 | 3,20 | 1452 | 3,24 | Win003_C02 | 1,33 |
| PU28528 | 3,18 | 1403 | 3,22 | Win020_C13 | 1,33 |
| PU25399 | 3,18 | 283 | 3,20 | Win015_C06 | 1,29 |
| PU23017 | 3,17 | 2923 | 3,20 | Win009_C02 | 1,29 |

***Table S3:*** *Identification of the forty most important variables for each block of the Hybrid Aspen data according to their relevance for the interpretation of the global variation (of the original OnPLS model described in Section 2). The global MB-VIOP values are provided in arbitrary units (a.u.).*

| THE 120 MOST IMPORTANT VARIABLES FOR THE LOCALLY JOINT VARIATION (40 VARIABLES PER BLOCK) | | | | | |
|---|---|---|---|---|---|
| Transcript variables | Local MB-VIOP values (a.u.) | Protein variables | Local MB-VIOP values (a.u.) | Metabolite variables | Local MB-VIOP values (a.u.) |
| PU06931 | 8,05 | 966 | 9,76 | Win031_C01 | 5,39 |
| PU07326 | 6,52 | 2121 | 8,23 | Win021_C05 | 4,67 |
| PU06434 | 6,21 | 1115 | 7,97 | Win034_C06 | 4,42 |
| PU03040 | 6,11 | 969 | 7,56 | Win034_C04 | 4,23 |
| PU01604 | 5,91 | 287 | 7,32 | Win022_C04 | 2,66 |
| PU08326 | 5,76 | 2805 | 6,84 | Win029_C03 | 2,45 |
| PU30269 | 5,57 | 2368 | 6,46 | Win018_C01 | 2,35 |
| PU08307 | 5,57 | 2364 | 6,45 | Win034_C02 | 2,34 |
| PU07241 | 5,55 | 2969 | 6,43 | Win004_C07 | 2,21 |
| PU07213 | 5,52 | 1991 | 6,17 | Win022_C05 | 2,18 |
| PU04361 | 5,43 | 164 | 5,91 | Win022_C03 | 2,06 |
| PU27830 | 5,41 | 3097 | 5,90 | Win029_C07 | 2,04 |
| PU31267 | 5,36 | 757 | 5,66 | Win013_C11 | 1,93 |
| PU07802 | 5,16 | 2119 | 5,52 | Win013_C08 | 1,87 |
| PU27837 | 5,01 | 121 | 5,33 | Win007_C09 | 1,66 |
| PU28081 | 4,86 | 439 | 5,31 | Win008_C08 | 1,62 |
| PU06797 | 4,82 | 527 | 5,23 | Win003_C04 | 1,58 |
| PU28220 | 4,69 | 795 | 5,18 | Win026_C07 | 1,55 |
| PU07004 | 4,67 | 986 | 5,12 | Win032_C01 | 1,50 |
| PU06101 | 4,62 | 751 | 4,89 | Win023_C09 | 1,47 |
| PU06985 | 4,61 | 2916 | 4,88 | Win008_C10 | 1,44 |
| PU07966 | 4,58 | 3061 | 4,82 | Win020_C14 | 1,44 |
| PU06614 | 4,58 | 718 | 4,64 | Win020_C12 | 1,36 |
| PU07280 | 4,53 | 1119 | 4,53 | Win026_C08 | 1,35 |
| PU30499 | 4,52 | 1960 | 4,44 | Win014_C02 | 1,33 |
| PU00630 | 4,50 | 1829 | 4,43 | Win016_C02 | 1,32 |
| PU08205 | 4,46 | 2124 | 4,35 | Win019_C03 | 1,30 |
| PU31286 | 4,44 | 2483 | 4,33 | Win015_C02 | 1,29 |
| PU08286 | 4,40 | 372 | 4,30 | Win034_C05 | 1,28 |
| PU06604 | 4,39 | 374 | 4,27 | Win007_C04 | 1,27 |
| PU22639 | 4,37 | 435 | 4,25 | Win020_C18 | 1,25 |
| PU00660 | 4,33 | 994 | 4,14 | Win028_C02 | 1,25 |
| PU03044 | 4,33 | 984 | 4,09 | Win027_C01 | 1,25 |
| PU05694 | 4,30 | 1259 | 4,01 | Win024_C07 | 1,24 |
| PU02114 | 4,25 | 1028 | 4,00 | Win004_C02 | 1,24 |
| PU04375 | 4,25 | 2839 | 3,97 | Win010_C03 | 1,23 |
| PU06213 | 4,21 | 625 | 3,93 | Win020_C11 | 1,22 |
| PU28218 | 4,08 | 1297 | 3,88 | Win026_C05 | 1,21 |
| PU30908 | 4,07 | 869 | 3,88 | Win018_C11 | 1,19 |
| PU08342 | 4,05 | 109 | 3,81 | Win006_C01 | 1,18 |

**Table S4:** *Identification of the forty most important variables for each block of the Hybrid Aspen data according to their relevance for the interpretation of the local variation (of the original OnPLS model described in Section 2). The local MB-VIOP values are provided in arbitrary units (a.u.).*

| THE 80 MOST IMPORTANT VARIABLES FOR THE UNIQUE VARIATION (40 VARIABLES PER BLOCK) | | | |
|---|---|---|---|
| Transcript variables | Unique MB-VIOP values (a.u.) | Metabolite variables | Unique MB-VIOP values (a.u.) |
| PU28218 | 6,05 | Win007_C09 | 5,83 |
| PU27903 | 5,92 | Win015_C08 | 4,48 |
| PU22619 | 5,64 | Win016_C04 | 4,25 |
| PU21598 | 5,60 | Win023_C09 | 3,56 |
| PU28093 | 5,43 | Win015_C06 | 3,24 |
| PU28695 | 5,38 | Win014_C02 | 3,16 |
| PU28785 | 5,19 | Win020_C06 | 2,87 |
| PU22718 | 4,74 | Win007_C03 | 2,61 |
| PU23246 | 4,58 | Win022_C02 | 2,19 |
| PU23171 | 4,48 | Win022_C12 | 2,10 |
| PU26833 | 4,39 | Win010_C16 | 2,07 |
| PU26977 | 4,32 | Win018_C10 | 2,02 |
| PU23160 | 4,31 | Win023_C03 | 1,91 |
| PU05336 | 4,25 | Win023_C07 | 1,90 |
| PU11583 | 4,13 | Win020_C12 | 1,89 |
| PU30650 | 4,09 | Win007_C04 | 1,87 |
| PU23500 | 4,04 | Win023_C05 | 1,85 |
| PU23219 | 4,00 | Win021_C05 | 1,85 |
| PU27204 | 3,96 | Win034_C03 | 1,83 |
| PU27020 | 3,93 | Win032_C02 | 1,80 |
| PU11487 | 3,93 | Win011_C02 | 1,76 |
| PU25908 | 3,91 | Win013_C03 | 1,75 |
| PU27299 | 3,91 | Win020_C10 | 1,71 |
| PU11448 | 3,90 | Win007_C10 | 1,70 |
| PU23792 | 3,90 | Win029_C03 | 1,67 |
| PU22279 | 3,90 | Win016_C02 | 1,59 |
| PU26882 | 3,87 | Win016_C05 | 1,56 |
| PU20096 | 3,85 | Win004_C07 | 1,55 |
| PU08307 | 3,84 | Win020_C09 | 1,54 |
| PU30561 | 3,84 | Win005_C05 | 1,49 |
| PU05084 | 3,81 | Win018_C06 | 1,46 |
| PU23192 | 3,80 | Win003_C06 | 1,45 |
| PU23215 | 3,80 | Win007_C08 | 1,34 |
| PU27202 | 3,80 | Win019_C03 | 1,31 |
| PU11653 | 3,76 | Win033_C01 | 1,30 |
| PU26956 | 3,75 | Win009_C10 | 1,29 |
| PU30655 | 3,74 | Win014_C08 | 1,27 |
| PU07802 | 3,74 | Win013_C09 | 1,27 |
| PU10587 | 3,73 | Win021_C03 | 1,25 |
| PU09082 | 3,71 | Win002_C03 | 1,24 |

***Table S5:*** *Identification of the forty most important variables for each block of the Hybrid Aspen data according to their relevance for the interpretation of the unique model components. The variables of the proteomics data block did not contribute to explain the unique variation of the original OnPLS model, hence, their values are not provided. The unique MB-VIOP values are provided in arbitrary units (a.u.).*

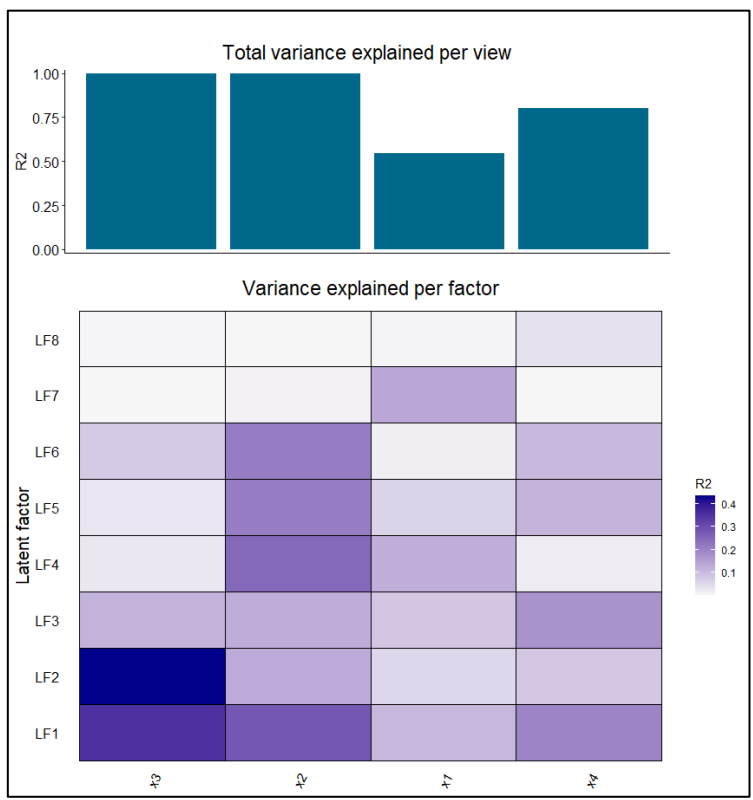

***Figure S1:*** *Total explained variance for the SD16_235GLU dataset from an 8-component MOFA model.*

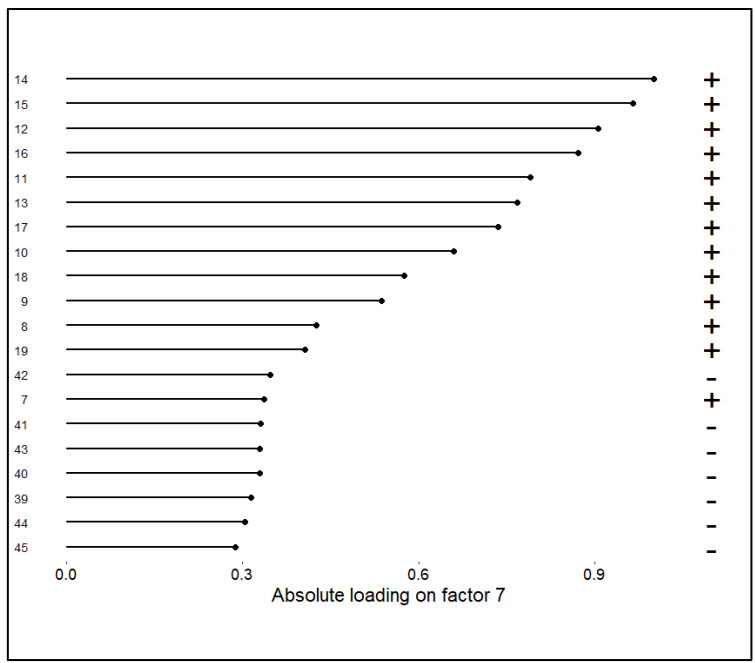

***Figure S2:*** *Absolute loading plot for the 7$^{th}$ component found by MOFA using the synthetic dataset.*

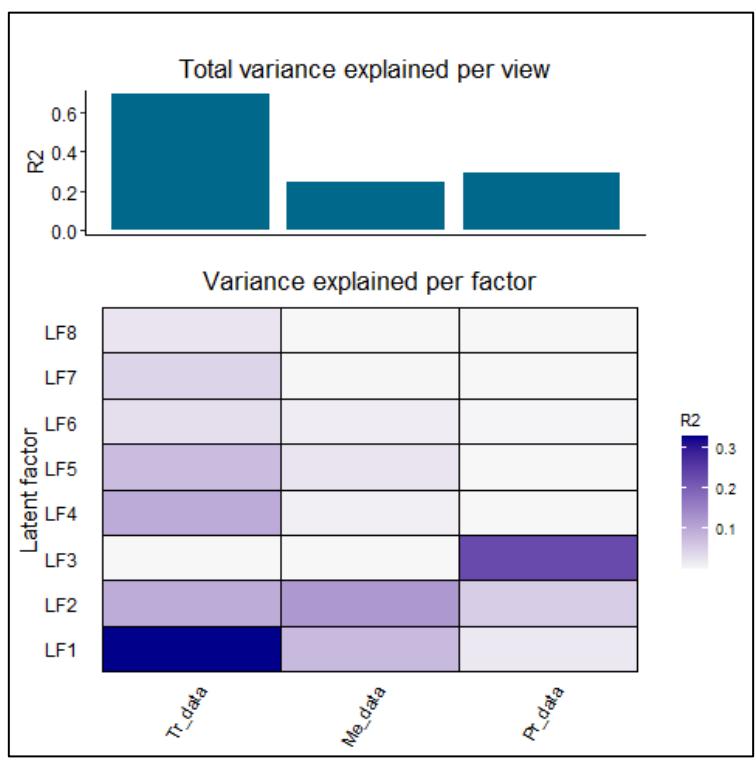

*Figure S3:* Total explained variance for the Hybrid Aspen dataset from an 8-component MOFA model.

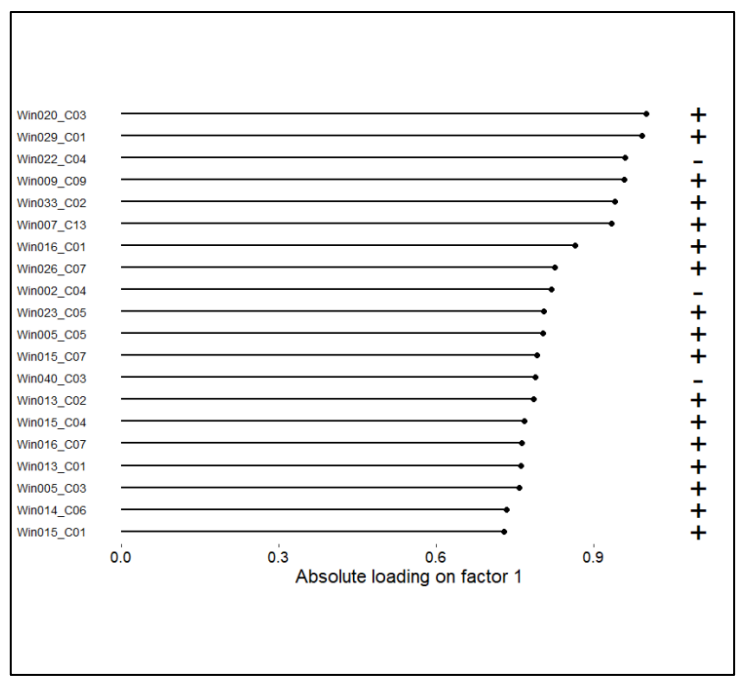

*Figure S4:* Absolute loading plot including metabolite variables for the 1st global component found by MOFA using the Hybrid Aspen dataset.

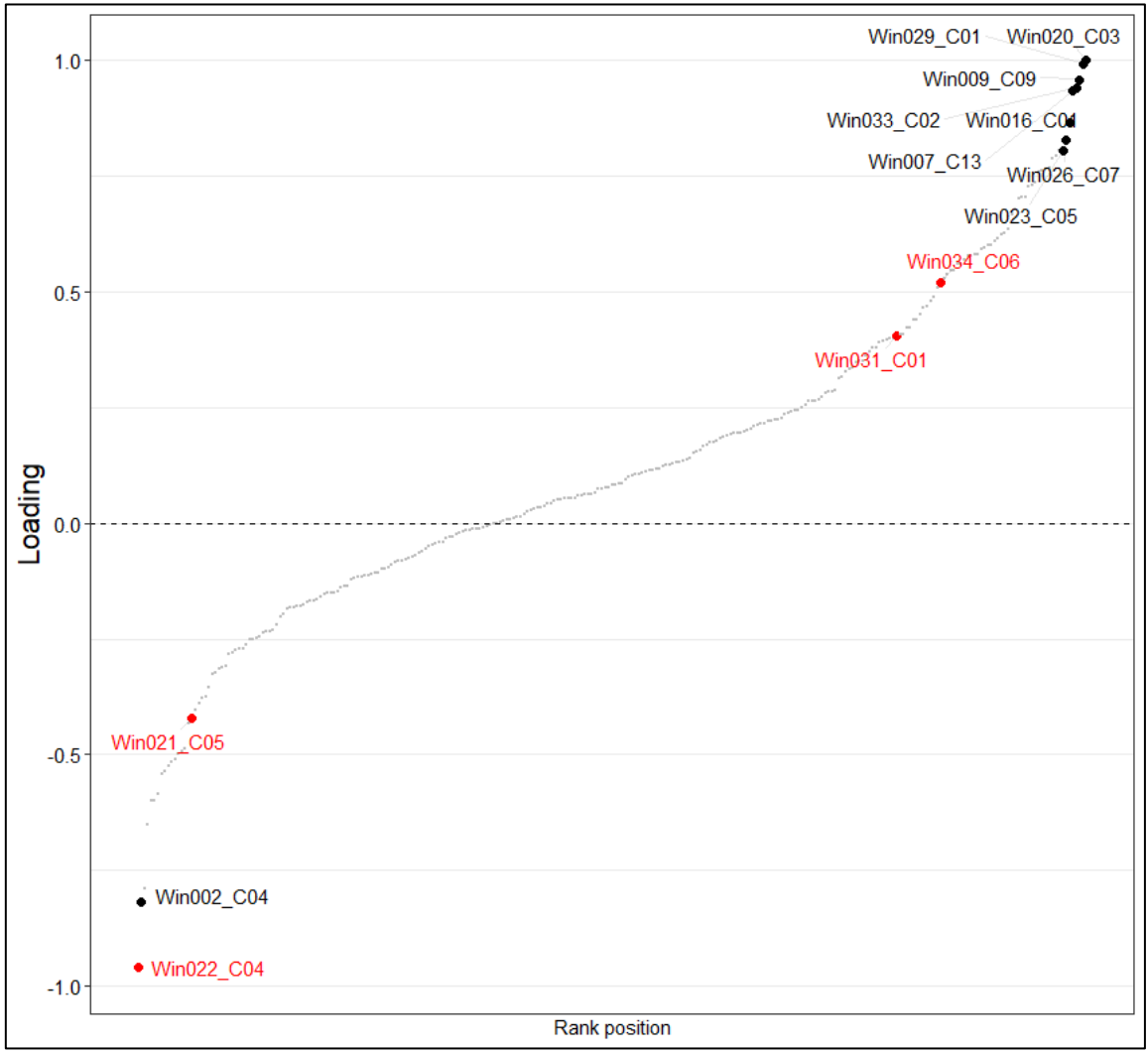

***Figure S5:*** *Loading vs Rank position plot highlighting some relevant metabolite variables for the 1$^{st}$ global component found by MOFA using the Hybrid Aspen dataset.*